\documentclass{aa}

\usepackage{natbib}
\usepackage{graphicx}
\usepackage{txfonts}
\usepackage{longtable}
\usepackage{appendix}
\usepackage{color}

\begin{document} 

   \title{A systematic study of silicate absorption features in heavily obscured AGNs observed by Spitzer/IRS}

  \author{T. Tsuchikawa, 
          \inst{1}
          H. Kaneda,\inst{1} S. Oyabu,\inst{2} T. Kokusho,\inst{1} H. Kobayashi,\inst{1} M. Yamagishi\inst{3} \and Y. Toba\inst{4,5,6} 
          }

   \institute{Graduate School of Science, Nagoya University, 
              Furo-cho, Chikusa-ku, Nagoya, Aichi 464-8602, Japan\\
              \email{tsuchikawa@u.phys.nagoya-u.ac.jp}
        \and
            Institute of Liberal Arts and Sciences, Tokushima University, 1-1 Minami-Jyosanjima, Tokushima-shi, Tokushima, 770-8502, Japan  
         \and
             Institute of Space and Astronautical Science, Japan Aerospace Exploration Agency, 3-1-1 Yoshinodai, Chuo-ku, Sagamihara, Kanagawa, 252-5210, Japan
	\and
		Department of Astronomy, Kyoto University, Kitashirakawa-Oiwake-cho, Sakyo-ku, Kyoto 606-8502, Japan
	\and
		Academia Sinica Institute of Astronomy and Astrophysics, 11F of Astronomy-Mathematics Building, AS/NTU, No.1, Section 4, Roosevelt Road, Taipei 10617, Taiwan
	\and
		Research Center for Space and Cosmic Evolution, Ehime University, 2-5 Bunkyo-cho, Matsuyama, Ehime 790-8577, Japan
             }

   \date{Received February 3, 2021; accepted May 10, 2021}

\abstract{Heavily obscured active galactic nuclei (AGNs) are known to show deep silicate absorption features in the mid-infrared (IR) wavelength range of 10--20~$\mu$m. The detailed profiles of the features reflect the properties of silicate dust, which are likely to include information on AGN activities obscured by large amounts of dust.}
{We reveal AGN activities obscured by large amounts of dust through the silicate dust properties obtained from the mid-IR spectral bands.}
{We select 115 mid-IR spectra of heavily obscured AGNs observed by Spitzer/IRS, and systematically analyze the composition of silicate dust by spectral fitting using the 10~$\mu$m amorphous and 23~$\mu$m crystalline bands.}
{We find that the main component of the silicate dust obscuring AGNs is amorphous olivine, the median mass column density of which is one order of magnitude higher than those of the minor components of amorphous pyroxene and crystalline forsterite. The median mass fraction of the amorphous pyroxene, $\sim$2\%, is significantly lower than that of the diffuse interstellar medium (ISM) dust in our Galaxy, while the median mass fraction of the crystalline forsterite, $\sim$6\%, is higher than that of the diffuse ISM dust. We also find that the mass fractions of the amorphous pyroxene and the crystalline forsterite positively correlate with each other.}
{The low mass fraction of the amorphous pyroxene suggests that the obscuring silicate dust is newly formed, originating from starburst activities. The relatively high mass fraction of crystalline forsterite implies that the silicate dust is processed in the high temperature environment close to the nucleus and transported to outer cooler regions by molecular outflows. The positive correlation between the mass fractions can be naturally explained considering that amorphous pyroxene is transformed from crystalline forsterite by ion bombardments. We also find that spectra with high ratios of the $\rm H_2O$ ice absorption to silicate mass column density tend to indicate low mass fractions of amorphous pyroxene and crystalline forsterite, which is consistent with the scenario of the thermal dust processing close to the nucleus.}

\titlerunning{Systematic study of silicate dust in heavily obscured AGNs}
\authorrunning{T. Tsuchikawa et al. }

   \maketitle

\section{Introduction}

The central engine of an active galactic nucleus (AGN) is considered to be surrounded by thick gas and dust, which have a toroidal structure predicted by the classical unified model \citep{Antonucci1993,Urry1995}. 
Near- to mid-infrared (IR) and hard X-ray observations over the last two decades have revealed the existence of a lot of galaxies harboring AGNs despite no optical signatures of AGN \citep[e.g., ][]{Imanishi2000,Risaliti2000}. The hidden AGNs are heavily obscured by such dense gas and dust that ionizing photons cannot escape from the nuclear region, and thus narrow line region (NLR), which is an ionized polar region emitting optical lines, is considered not to be developed \citep[e.g.,][]{Imanishi2006}. The standard supermassive black hole (SMBH) and galaxy co-evolutionary scenario predicts that heavily obscured AGNs are in an early phase of the SMBH growth \citep{Hopkins2008}. Therefore, in order to understand the AGN evolution, it is important to study the nature of the heavily obscured AGN activities and the obscuring material.

	Amorphous silicate is a major component of the interstellar dust \citep{Mathis1977}, the band features of which peak at the mid-IR wavelengths of ${\sim}10$ and ${\sim}18$~$\mu$m in Si-O stretching and O-Si-O bending modes, respectively.
    These bands with the absolute value of the silicate strength > 1 are seen in 97 out of 196 mid-IR spectra of nearby AGNs and ultra-luminous IR galaxies (ULIRGs) \citep{Hao2007}.
    In particular, owing to the large samples observed by Spitzer/IRS \citep{Houck2004}, mid-IR properties of low-redshift AGNs ($z<1$) have been studied well \citep[e.g.,][]{Weedman2005, Hao2007, Imanishi2007, Spoon2007}. In these studies, the strengths
    of the silicate features at the peak wavelengths were often analyzed, from which it was found that the silicate features in QSOs or type-1 AGNs are observed in emission or weak absorption, while those in type-2 AGNs only in absorption \citep[e.g.,][]{Hao2007,Spoon2007,Hatziminaoglou2015}.
   The silicate features in heavily obscured AGNs tend to show deep absorption, the optical depths of which are greater than 2 \citep[e.g., ][]{Hao2007, Imanishi2007}.
    In general, the strengths of the silicate features depend on the geometrical structure of obscuring clouds along the line of sight, while their profile shapes reflect the dust properties, such as chemical composition and crystallinity \citep{Henning2010}.
    For example, the peaks of the 10~${\mu}$m silicate emission features in the spectra of QSOs and type-1 AGNs are known to shift toward longer wavelengths than that of typical amorphous silicate, which is explained by higher porosity \citep{Li2008}, a larger size of dust \citep{Smith2010}, or other dust components \citep{Kemper2007, Srinivasan2017}.

    In the Circinus galaxy, a famous type-2 AGN, crystalline silicate is detected with ground-based telescopes, showing that the crystallinity is similar to that of the diffuse interstellar medium (ISM) silicate in our Galaxy \citep{DoDuy2019}.
    On the other hand, \citet{Spoon2006} find that 12 ULIRGs with deep silicate absorption features in the mid-IR spectra have silicate dust with crystallinity higher than the diffuse ISM silicate dust in our Galaxy. 
    \citet{Spoon2006} suggest that the high crystallinity originates from starburst activities, while \citet{Kemper2011} predict that the high crystallinity nature needs an additional source producing crystalline silicate other than starburst activities.
  
    \citet{Tsuchikawa2019} investigate the profile shapes of the 10~$\mu$m silicate features in 39 heavily obscured AGNs observed by AKARI/IRC and Spizter/IRS, and find that the profile shapes are notably similar as a whole. Nevertheless, they also find that the profile shapes are different in the wing on the shorter wavelength side, which indicates the variation of the mineralogical composition of silicate dust from galaxy to galaxy.
In this paper, we systematically quantify the detailed differences in the silicate features in heavily obscured AGNs, and determine the dust properties for each galaxy. Using the dust properties thus obtained, we aim to reveal obscured AGN activities and the surrounding environments.

\section{Sample selection}
We use mid-IR spectra observed by the Infrared Spectrograph \citep[IRS; ][]{Houck2004} aboard the Spitzer Space Telescope \citep{Werner2004}. IRS low-resolution (LR) spectra were observed using two spectroscopic modules (Short-Low; SL, Long-Low; LL), covering the wavelength ranges of 5.2--14.5 and 14.0--38.0~$\mu$m, and therefore we can study both the silicate 10 and 18~$\mu$m features for low-redshift galaxies.
The spectroscopic data were retrieved from the Cornell AtlaS of Spitzer/IRS Sources \citep[CASSIS; ][]{Lebouteiller2011, Lebouteiller2015} version LR7 which contain more than 10,000 LR spectra.
The CASSIS provides the spectra reduced by the two spectral extraction techniques which are the optimal extraction for point-like sources and the tapered column extraction for partially-extended sources. 
We selected either of the two spectral extraction methods for each spectrum according to the pipeline message based on the source spatial extent.  
Most of the spectra retrieved from CASSIS show discontinuity in the flux density between the SL and LL spectra, which is mainly caused by difference in the extraction aperture and slit width between the SL and LL modules, and thus we performed a spectral stitching by scaling the SL spectra.

Among the spectra of extragalactic sources in CASSIS, which we searched for according to the science category of the approved Spitzer programs, we retrieved those with apparently deep silicate absorption features.
We selected 115 obscured AGNs using the following three selection criteria from the sample retrieved from CASSIS: (1) the apparent peak optical depth of the 10~$\mu$m silicate feature is larger than 1.5. The apparent peak optical depth is defined by $-{\rm ln}({f_{\rm obs}(10~\mu{\rm m})}/{f_{\rm cont}(10~\mu{\rm m})})$, where $f_{\rm obs}(10~\mu{\rm m})$ and $f_{\rm cont}(10~\mu{\rm m})$ are an observed flux density and absorption-free continuum flux density, respectively, both at the peak wavelength of the 10~$\mu$m silicate feature \citep{Spoon2007}. The absorption-free continuum is assumed to be the power-law function determined from the flux densities at 7.1 and 14.2$\,{\rm {\mu}m}$, following the method defined by e.g., \citet{Imanishi2007}.
Since some spectra showed low signal-to-noise ratios ($S/N$) at the bottom of the 10~$\mu$m silicate feature, we performed spectral binning so that $S/N$ at 10~$\mu$m exceeded 5, and then calculated the apparent optical depth of the 10~$\mu$m silicate feature at the peak wavelength.
(2) The equivalent width of the polycyclic aromatic hydrocarbon (PAH) 6.2~$\mu$m feature is smaller than 270~nm, by which we can select AGN-dominated galaxies \citep{Stierwalt2013}. We calculated the equivalent width by integrating the flux densities above the power-law local continuum determined from the flux densities at 6.0 and 6.45$\,{\rm {\mu}m}$ \citep[e.g., ][]{Imanishi2007}. 
By this criterion, we analyze the silicate features robustly with less contamination of the PAH emission. Fifty one sources which meet criteria (1) and (3) are removed out of 166 based on criterion (2). 
(3) The redshift $z$ is lower than 0.35. The spectra of low-redshift galaxies selected by the criterion cover both silicate 10 and 18~$\mu$m features.
Table~\ref{table:sample} summarizes the general properties of the 115 obscured AGNs selected by the above three criteria.  
Although IRAS~13454--2956N, WISEA~142500.11+325949.9 and WISEA~J231813.00--004125.9 meet all the three selection criteria, we excluded them from our sample because the spectral data of one of the orders in the LR module are not available. IRAS~16255+2801 is confused with a planetary nebula in our Galaxy, and hence also excluded from our sample.

\begin{table*}
\caption{General properties of our sample}             
\label{table:sample}     
\centering                          
 \begin{tabular}{l c c c c}        
\hline\hline                 
 Name & AORkey & R.A. (J2000) & Dec. (J2000) & $z$ \\    
 (1)&(2)&(3)&(4)&(5)\\
\hline

IRAS~00091-0738 & 10440960, 10108928 & 00h11m43.2s & 	--07d22m06s & 0.1184 \\
IRAS~F00183-7111 & 7556352 & 00h20m34.6s & 	--70d55m26s & 0.3270 \\
IRAS~00188-0856 & 4962560 & 00h21m26.4s & 	--08d39m27s & 0.1284 \\
IRAS~00397-1312 & 4963584 & 00h42m15.4s & 	--12d56m03s & 0.2617 \\
IRAS~00406-3127 & 4964096 & 00h43m03.1s & 	--31d10m49s & 0.3424 \\
IRAS~01166-0844SE & 10441984, 10109952 & 01h19m07.8s & 	--08d29m12s & 0.1180 \\
IRAS~F01173+1405 & 20356352 & 01h20m02.6s & 	+14d21m42s & 0.0312 \\
IRAS~F01197+0044 & 22132224 & 01h22m18.1s & 	+01d00m25s & 0.0555 \\
IRAS~01199-2307 & 4964864 & 01h22m20.8s & 	--22d51m57s & 0.1562 \\
IRAS~01298-0744 & 4965120 & 01h32m21.4s & 	--07d29m08s & 0.1362 \\
IRAS~01355-1814 & 4965376 & 01h37m57.4s & 	--17d59m20s & 0.1920 \\
IRAS~F01478+1254 & 23012864 & 01h50m28.4s & 	+13d08m58s & 0.1470 \\
IRAS~01569-2939 & 10110208 & 01h59m13.7s & 	--29d24m34s & 0.1400 \\
IRAS~02438+2122 & 10508544, 20353792 & 02h46m39.1s & 	+21d35m10s & 0.0233 \\
IRAS~02455-2220 & 4967680 & 02h47m51.2s & 	--22d07m38s & 0.2840 \\
IRAS~02530+0211 & 6652160 & 02h55m34.4s & 	+02d23m41s & 0.0276 \\
IRAS~03158+4227 & 12256256 & 03h19m11.9s & 	+42d38m25s & 0.1344 \\
$\rm NGC~1377^{a}$ & 9511424 & 03h36m40.1s & 	--20d54m02s & 0.0060 \\
IRAS~03538-6432 & 4968192 & 03h54m25.2s & 	--64d23m44s & 0.3007 \\
IRAS~03582+6012 & 20341504 & 04h02m32.9s & 	+60d20m41s & 0.0300 \\
IRAS~04074-2801 & 25185536 & 04h09m30.4s & 	--27d53m43s & 0.1537 \\
IRAS~04313-1649 & 4968960 & 04h33m37.0s & 	--16d43m31s & 0.2680 \\
IRAS~04384-4848 & 6650880 & 04h39m50.8s & 	--48d43m17s & 0.2035 \\
ESO~203-IG001 & 20334080 & 04h46m49.5s & 	--48d33m30s & 0.0529 \\
IRAS~05020-2941 & 25185792 & 05h04m00.7s & 	--29d36m54s & 0.1544 \\
IRAS~F06076-2139 & 20359680 & 06h09m45.7s & 	--21d40m24s & 0.0374 \\
IRAS~06206-6315 & 4969984 & 06h21m00.8s & 	--63d17m23s & 0.0924 \\
IRAS~06301-7934 & 4970240 & 06h26m42.2s & 	--79d36m30s & 0.1564 \\
IRAS~06361-6217 & 4970496 & 06h36m35.7s & 	--62d20m31s & 0.1596 \\
IRAS~F07224+3003 & 19165184 & 07h25m37.2s & 	+29d57m14s & 0.0188 \\
IRAS~07251-0248 & 20346112 & 07h27m37.6s & 	--02d54m54s & 0.0876 \\
MCG~+02-20-003 & 20353280 & 07h35m43.4s & 	+11d42m34s & 0.0163 \\
SDSS~J082001.72+505039.1 & 23014400 & 08h20m01.7s & 	+50d50m39s & 0.2173 \\
IRAS~08201+2801 & 18202112 & 08h23m12.6s & 	+27d51m40s & 0.1678 \\
IRAS~F08520-6850 & 20343808 & 08h52m32.0s & 	--69d01m54s & 0.0451 \\
IRAS~08572+3915 & 4972032 & 09h00m25.3s & 	+39d03m54s & 0.0584 \\
IRAS~09039+0503 & 10443776, 10104064 & 09h06m34.0s & 	+04d51m25s & 0.1251 \\
IRAS~09539+0857 & 10444032, 11676160 & 09h56m34.3s & 	+08d43m05s & 0.1289 \\
IRAS~F10038-3338 & 20352256 & 10h06m04.6s & 	--33d53m06s & 0.0342 \\
IRAS~10091+4704 & 4973824 & 10h12m16.7s & 	+46d49m42s & 0.2460 \\
IRAS~F10112-0040 & 15069440 & 10h13m46.8s & 	-00d54m51s & 0.0425 \\
IRAS~10173+0828 & 14838528, 20314880 & 10h20m00.2s & 	+08d13m34s & 0.0491 \\
IRAS~F10237+4720 & 22117632 & 10h26m48.2s & 	+47d05m07s & 0.0589 \\
IRAS~10378+1109 & 4974336 & 10h40m29.1s & 	+10d53m17s & 0.1363 \\
IRAS~10485-1447 & 10444800, 10105088 & 10h51m03.0s & 	--15d03m22s & 0.1330 \\
IRAS~11028+3130 & 18203392 & 11h05m37.5s & 	+31d14m31s & 0.1986 \\
IRAS~11038+3217 & 4975104 & 11h06m35.7s & 	+32d01m46s & 0.1300 \\
IRAS~11095-0238 & 4975360 & 11h12m03.3s & 	--02d54m24s & 0.1066 \\
IRAS~11130-2659 & 10105600 & 11h15m31.5s & 	--27d16m22s & 0.1361 \\
IRAS~11180+1623 & 18203648 & 11h20m41.7s & 	+16d06m56s & 0.1660 \\
IRAS~11223-1244 & 4976128 & 11h24m50.7s & 	--13d01m16s & 0.1990 \\
IRAS~11506+1331 & 10445312, 10111488 & 11h53m14.1s & 	+13d14m26s & 0.1273 \\
IRAS~11524+1058 & 18203904 & 11h55m05.1s & 	+10d41m22s & 0.1787 \\
IRAS~11582+3020 & 4976384 & 12h00m46.8s & 	+30d04m14s & 0.2230 \\
IRAS~12032+1707 & 4976896 & 12h05m47.7s & 	+16d51m08s & 0.2178 \\
IRAS~12127-1412 & 10445824, 10105856 & 12h15m19.1s & 	--14d29m41s & 0.1330 \\
IRAS~F12224-0624 & 20367104 & 12h25m03.9s & 	--06d40m52s & 0.0264 \\
NGC~4418 & 4935168 & 12h26m54.6s & 	-00d52m40s & 0.0073 \\
IRAS~12359-0725 & 10106112 & 12h38m31.6s & 	--07d42m25s & 0.1380 \\
IRAS~12447+3721 & 25187840 & 12h47m07.7s & 	+37d05m36s & 0.1580 \\
IRAS~F13045+2354 & 4168448 & 13h07m00.6s & 	+23d38m04s & 0.2750 \\
\hline

\end{tabular}

\end{table*}

\setcounter{table}{0}
\begin{table*}
\caption{Continued.}             
\centering     
 \begin{tabular}{l c c c c}        
\hline\hline                 
 Name & AORkey & R.A. (J2000) & Dec. (J2000) & $z$ \\    
 (1)&(2)&(3)&(4)&(5)\\
\hline

IRAS~13106-0922 & 25186048 & 13h13m14.6s & 	--09d38m08s & 0.1745 \\
IRAS~F13279+3401 & 12235264 & 13h30m15.2s & 	+33d46m29s & 0.0230 \\
IRAS~13352+6402 & 4979968 & 13h36m51.1s & 	+63d47m04s & 0.2366 \\
Mrk~273 & 4980224 & 13h44m42.1s & 	+55d53m13s & 0.0378 \\
IRAS~14070+0525 & 4980992 & 14h09m31.2s & 	+05d11m31s & 0.2644 \\
IRAS~14121-0126 & 25186304 & 14h14m45.5s & 	--01d40m55s & 0.1502 \\
IRAS~F14242+3258 & 14084352 & 14h26m23.8s & 	+32d44m35s & 0.1760 \\
IRAS~14348-1447 & 4981248 & 14h37m38.2s & 	--15d00m24s & 0.0830 \\
IRAS~F14394+5332 & 29040128 & 14h41m04.3s & 	+53d20m08s & 0.1045 \\
IRAS~F14511+1406 & 4168960 & 14h53m31.5s & 	+13d53m58s & 0.1390 \\
IRAS~F14554+3858 & 28244224 & 14h57m22.7s & 	+38d46m28s & 0.0735 \\
IRAS~15225+2350 & 10112512 & 15h24m43.9s & 	+23d40m10s & 0.1390 \\
IRAS~15250+3609 & 4983040 & 15h26m59.3s & 	+35d58m37s & 0.0552 \\
Arp~220 & 4983808 & 15h34m57.2s & 	+23d30m11s & 0.0181 \\
FESS~J160655.82+541500.7 & 24189952 & 16h06m55.8s & 	+54d15m00s & 0.2060 \\
IRAS~F16073+0209 & 17546496 & 16h09m49.7s & 	+02d01m30s & 0.2230 \\
IRAS~16090-0139 & 4984576 & 16h11m40.4s & 	--01d47m05s & 0.1336 \\
FESS~J161759.22+541501.3 & 24163328 & 16h17m59.2s & 	+54d15m01s & 0.1339 \\
IRAS~F16156+0146 & 17546752 & 16h18m09.3s & 	+01d39m22s & 0.1320 \\
IRAS~F16242+2218 & 17547008 & 16h26m26.0s & 	+22d11m45s & 0.1570 \\
IRAS~F16305+4823 & 22135040 & 16h31m58.7s & 	+48d17m22s & 0.0874 \\
IRAS~16300+1558 & 4985088 & 16h32m21.4s & 	+15d51m45s & 0.2417 \\
IRAS~16455+4553 & 14875136 & 16h46m58.9s & 	+45d48m22s & 0.1906 \\
IRAS~16468+5200W & 10107136 & 16h48m01.3s & 	+51d55m43s & 0.1500 \\
IRAS~16468+5200E & 10106880 & 16h48m01.6s & 	+51d55m44s & 0.1500 \\
NGC~6240 & 4985600 & 16h52m58.8s & 	+02d24m03s & 0.0245 \\
IRAS~17044+6720 & 10107904 & 17h04m28.4s & 	+67d16m28s & 0.1349 \\
IRAS~F17028+3616 & 27194112 & 17h04m33.5s & 	+36d12m18s & 0.0851 \\
IRAS~17068+4027 & 4986112 & 17h08m32.1s & 	+40d23m28s & 0.1790 \\
IRAS~17208-0014 & 4986624 & 17h23m21.9s & 	-00d17m00s & 0.0428 \\
IRAS~17463+5806 & 4987392 & 17h47m04.7s & 	+58d05m22s & 0.3090 \\
IRAS~17540+2935 & 18204928 & 17h55m56.1s & 	+29d35m26s & 0.1081 \\
IRAS~18443+7433 & 4987904 & 18h42m54.7s & 	+74d36m21s & 0.1347 \\
IRAS~18531-4616 & 4988160 & 18h56m53.0s & 	--46d12m46s & 0.1408 \\
IRAS~18588+3517 & 18205440 & 19h00m41.1s & 	+35d21m27s & 0.1067 \\
IRAS~20087-0308 & 4989440 & 20h11m23.8s & 	--02d59m50s & 0.1057 \\
IRAS~20100-4156 & 4989696 & 20h13m29.8s & 	--41d47m34s & 0.1296 \\
IRAS~20109-3003 & 14875904 & 20h14m05.5s & 	--29d53m53s & 0.1407 \\
IRAS~20286+1846 & 18205696 & 20h30m54.4s & 	+18d56m37s & 0.1358 \\
IRAS~20551-4250 & 4990208 & 20h58m26.7s & 	--42d39m01s & 0.0430 \\
IRAS~21077+3358 & 18205952 & 21h09m50.6s & 	+34d10m34s & 0.1767 \\
IRAS~21272+2514 & 4990464 & 21h29m29.3s & 	+25d27m55s & 0.1508 \\
IRAS~F21329-2346 & 10448640, 10108160 & 21h35m45.8s & 	--23d32m34s & 0.1251 \\
IRAS~F21541-0800 & 27441408 & 21h56m49.5s & 	--07d45m32s & 0.0551 \\
NGC~7172 & 18513920 & 22h02m01.9s & 	--31d52m11s & 0.0087 \\
IRAS~22088-1831W & 25189120 & 22h11m33.7s & 	--18d17m06s & 0.1702 \\
IRAS~22088-1831E & 25189376 & 22h11m33.8s & 	--18d17m05s & 0.1702 \\
IRAS~22116+0437 & 18206464 & 22h14m10.3s & 	+04d52m26s & 0.1938 \\
$\rm NGC~7479^{b}$ & 22093312 & 23h04m56.6s & 	+12d19m22s & 0.0079 \\
IRAS~23129+2548 & 4991488 & 23h15m21.4s & 	+26d04m32s & 0.1789 \\
IRAS~F23234+0946 & 10449152, 10108416 & 23h25m56.2s & 	+10d02m50s & 0.1279 \\
IRAS~23230-6926 & 4992000 & 23h26m03.5s & 	--69d10m20s & 0.1066 \\
IRAS~23253-5415 & 4992256 & 23h28m06.1s & 	--53d58m30s & 0.1300 \\
IRAS~23365+3604 & 4992512 & 23h39m01.2s & 	+36d21m09s & 0.0645 \\
\hline

\end{tabular}

\tablefoot{Column 1: the name of the object; Column 2: AORkey (Spitzer/IRS identification number); Columns 3, 4: the position of the object; Column 5: the redshift cited from the NASA/IPAC Extragalactic Database (NED).
\tablefoottext{a}{The spectrum of NGC~1377 is unavailable in CASSIS. We retrieved the spectral data from the summary of the SINGS Legacy project in the NASA/IPAC IR Science Archive (IRSA).}
\tablefoottext{b}{The SL order 2 spectrum of NGC~7479 is unavailable in CASSIS. We retrieved it from the Spitzer Heritage Archive (SHA).}
}
\end{table*}

\section{Spectral decomposition of the 10~$\mu$m silicate feature}
\subsection{Mid-IR spectral modeling}

   \begin{figure*}
	\centering
	\includegraphics[width=14cm,clip]{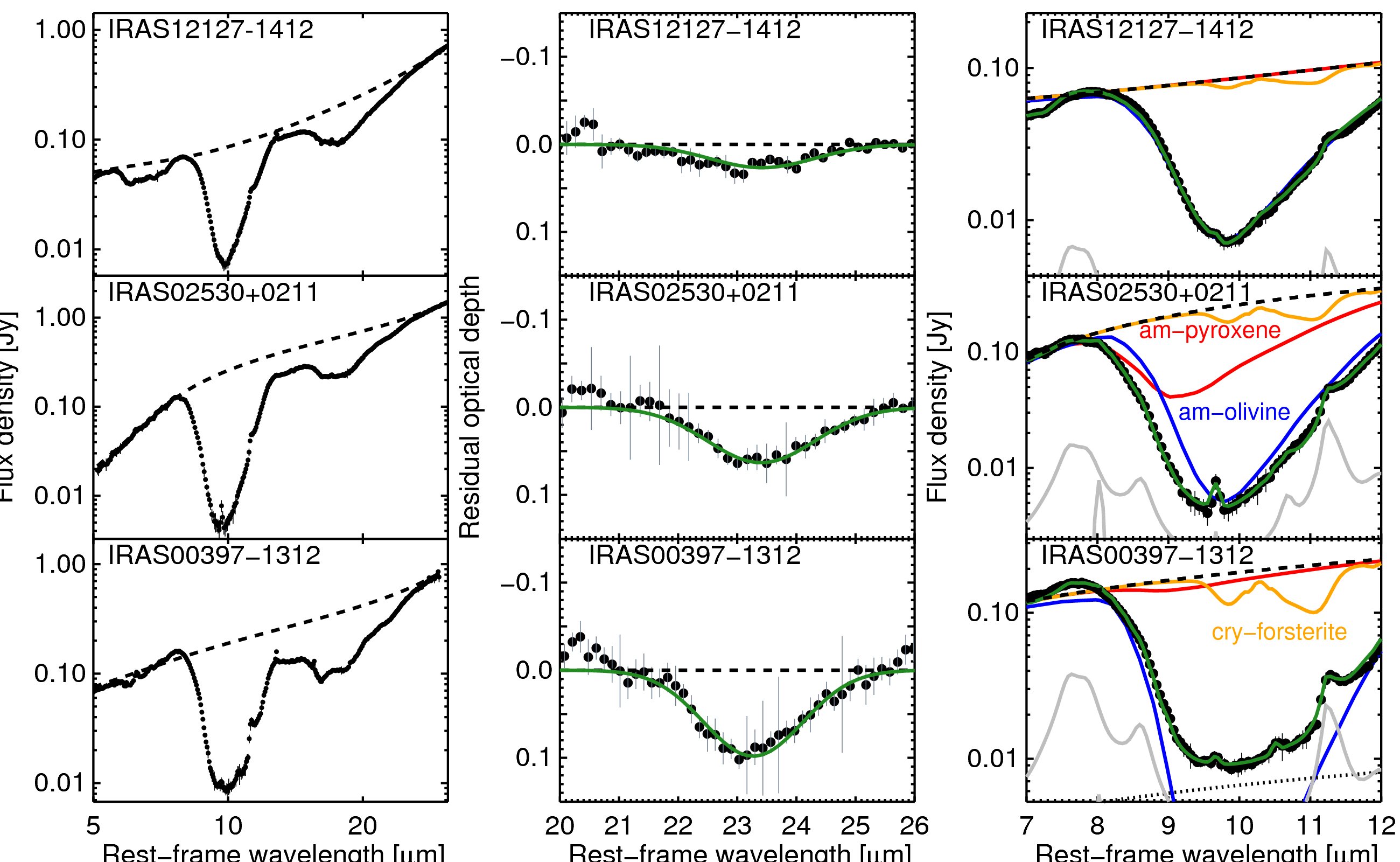}
	\caption{Examples of the 5.3--12~$\mu$m spectral fitting. Left panels show the retrieved mid-IR spectra and the estimated absorption-free continua. Middle panels show the Gaussian fits to the 23~$\mu$m features. Results of the 5.3--12~$\mu$m spectral fitting are shown in the right panels. In the right panels, the green solid lines show the best-fit models, the absorption components of which are composed of amorphous pyroxene (red lines; $F_{\rm hot, AGN}~{\rm exp}(-\tau_{\rm pyr})$ in equation 1), amorphous olivine (blue lines; $F_{\rm hot, AGN}~{\rm exp}(-\tau_{\rm ol})$) and crystalline forsterite (orange lines; $F_{\rm hot, AGN}~{\rm exp}(-\tau_{\rm cry})$). The black dashed, dotted and grey solid lines represent the absorption-free continuum ($F_{\rm hot, AGN}$), the unobscured continuum ($F_{\rm hot, unobs}$) and the PAH and line emission ($F_{\rm PAH} + F_{\rm line}$) components, respectively. Note that the unobscured continuum component falls below the plotted area for IRAS~12127--1412 and IRAS~02530+0211.}
	\label{fig:fit_result}%
   \end{figure*}
   
In order to investigate the dust properties, we decompose the 10~$\mu$m silicate feature in our sample by spectral fitting in the wavelength range of 5.3--12~$\mu$m, where we assume the screen geometry for simplicity. 
The fitting function, $F_{\nu}$, is described as
   \begin{equation}
    F_{\nu} = F_{\rm hot, AGN}~{\rm exp}(-\tau_{\rm sil}-\tau_{\rm ice+HAC}) + F_{\rm PAH} + F_{\rm line} + F_{\rm hot, unobs},
   \label{eq:model}
   \end{equation}
where $F_{\rm hot, AGN}$, $F_{\rm PAH}$, $F_{\rm line}$ and $F_{\rm hot, unobs}$ are the absorption-free continuum emission due to the hot dust heated by AGN, PAH and line emissions and an additional continuum emission due to unobscured hot dust, respectively. $\tau_{\rm sil}$ and $\tau_{\rm ice+HAC}$ are the optical depths of silicate dust and water ice plus hydrogenated amorphous carbon (HAC).

In the spectral fitting, we have to accurately determine $F_{\rm hot, AGN}$ to extract the properties of silicate dust. 
\citet{Tsuchikawa2019} reproduced the 2.5--12.5~$\mu$m spectrum of the heavily obscured AGN LEDA~1712304 well by a single-temperature blackbody emission as an absorption-free continuum, while the spectra in our sample are expected to be poorly fitted by the same model as used in \citet{Tsuchikawa2019} due to the variety of the spectra. 
\citet{Spoon2006} adopted the spline continuum using the pivot points anchored at the wavelengths of 5.6, 7.1~$\mu$m and the longer wavelength end of the spectra, and, for feature-less spectra, replaced the anchor point at 7.1~$\mu$m by the continuum at 7.9--8.0~$\mu$m. The method can be considered as good estimates for feature-less spectra as shown in figure~1 of \citet{Spoon2006}, while it does not match the spectra with other prominent spectral features such as those due to PAHs and ices in the wavelength range of 5--8~$\mu$m. 
\citet{Nardini2010} report that the 5--8~$\mu$m spectra of 164 ULIRGs, covering 68 spectra in our sample, are reproduced well using a power-law function as an absorption-free continuum. 
Therefore we reproduce the absorption-free continuum using a power-law function at 5.3--8~$\mu$m and a spline curve smoothly connected to the longer wavelength end of the spectrum. The spline curve is determined by using the pivot points anchored at 7.1 and 7.5~$\mu$m on the power-law continuum and 30 and 31~$\mu$m on the observed spectrum. 
The dashed lines in the left panel of Fig.~\ref{fig:fit_result} show three examples of the absorption-free continua obtained by the 5.3--12~$\mu$m spectral fitting.
The sample objects whose redshifts are larger than 0.3 are not observed at the rest wavelength range longer than 30~$\mu$m, and thus the pivot points at 30 and 31~$\mu$m are extrapolated from the observed spectra assuming a power-law function.

The optical depth of the 10~$\mu$m silicate feature, $\tau_{\rm sil}$, is composed of three dust species using the following equation:
   \begin{equation}
    \tau_{\rm sil} = \sum_i \tau_i = \sum_i N_i\kappa_i,
   \label{eq:eff_tau}
   \end{equation}
where $N_i$ and $\kappa_i$ are the mass column density and the mass absorption coefficient (MAC) of each silicate dust component, respectively.
We reproduce the absorption features due to water ice at 6~$\mu$m and hydrogenated amorphous carbon (HAC) at 6.85 and 7.25~$\mu$m considering a screen geometry that the hot dust in the nuclear region is covered with colder dust clouds with water ice and HAC. For water ice and HAC opacity models, we use the template derived from the spectrum of the deeply obscured ULIRG IRAS~F00183--7111 in \citet{Marshall2007} and a Gaussian function, respectively. However, the water ice template cannot well reproduce some of the spectra in our sample well, as also pointed out in \citet{Nardini2008}. For example, the spectrum of IRAS~12127--1412 shows a broader absorption feature, which extends to $\sim$7.5~$\mu$m, than that of IRAS~F00183--7111. The difference in the absorption feature is likely to be caused by complex compositions of processed ices including other organic species \citep{Boogert2015}. Thus we add other templates of water ice as additional absorption components derived from the spectra of IRAS~12127--1412 and NGC~4418, which show little PAH emission similar to that of IRAS~F00183--7111, by spline interpolation to estimate an absorption-free continuum. For the HAC absorptions, we fix the full width at half-maxima (FWHMs) at 0.17~$\mu$m, and the central wavelengths at 6.85 and 7.25~$\mu$m.
We apply the \citet{Draine2007} model to the PAH emission, $F_{\rm PAH}$, assuming the typical size distribution of the diffuse ISM of star-forming galaxies. The ionization fraction of PAHs is set to be free considering different radiation fields in the circumnuclear environments, as mentioned in \citet{Kaneda2008} for elliptical galaxies. 
We find a emission feature at 10.68~$\mu$m stronger than those predicted by the \citet{Draine2007} model, which is suggested to originate from dehydrogenated PAHs \citep{Mackie2015}. Hence we consider an additional emission component using a Drude function, both the FWHM and central wavelength are fixed according to the resonance parameter shown in \citet{Draine2007}. We also consider the interstellar extinction on the PAH emission assuming the well-mixed geometry, where we adopt the extinction curve described in \citet{Chiar2006}.
A Gaussian function is assumed for the profiles of atomic and molecular line emission components, $F_{\rm line}$, both widths and central wavelengths of which are fixed at the typical parameters described in \citet{Marshall2007}. 
\citet{Tsuchikawa2019} also suggest that the sharpness of the 10~$\mu$m silicate feature of heavily obscured AGNs is variable from galaxy to galaxy, which is likely to be attributed to the saturation by an unobscured emission component. 
Therefore, we added a continuum emission component of unobscured hot dust, $F_{\rm hot, unobs}$, using the spectral profile of $F_{\rm hot, AGN}$ in order to explain the difference in the sharpness\footnote{We verify the validity of the continuum emission component of unobscured hot dust in Appendix~\ref{sect:sharp}.}.

\subsection{Dust properties}

We decompose $\tau_{\rm sil}$ to three dust components, amorphous olivine ($\rm Mg_{2x}Fe_{2(1-x)}SiO_4$), amorphous pyroxene ($\rm Mg_{x}Fe_{1-x}SiO_3$) and crystalline forsterite ($\rm Mg_{2}SiO_4$). Amorphous olivine, the main component of the diffuse ISM in our Galaxy \citep{Draine2003}, is reported to reproduce the 10~$\mu$m feature of heavily obscured AGNs well \citep{Spoon2006, Tsuchikawa2019}. Crystalline forsterite is also known to compose the silicate dust in heavily obscured AGNs on the basis of detections of absorption features at 11, 16, 19 and 23~$\mu$m \citep{Spoon2006, Stierwalt2014}.
\citet{Tsuchikawa2019} report that the wings of the 10~$\mu$m silicate absorption features shown in the spectra of the heavily obscured AGNs significantly vary from galaxy to galaxy especially on the shorter wavelength side. 
In other words, the central wavelengths of the 10~$\mu$m features are variable among the heavily obscured AGNs.
The difference is likely to be caused by a mineralogical composition ratio of amorphous olivine and pyroxene, since the absorption efficiencies of amorphous olivine and pyroxene have different peak wavelengths of $\sim$9.8 and $\sim$9.3~$\mu$m, respectively \citep{Dorschner1995}. 
Accordingly, we also consider an amorphous pyroxene component in the spectral fitting. 

 The MACs of amorphous olivine and amorphous pyroxene, $\kappa_{\rm ol}$ and $\kappa_{\rm pyr}$, are represented by $3Q_{\rm abs, ol}/4a\rho_{\rm ol}$ and $3Q_{\rm abs, pyr}/4a\rho_{\rm pyr}$, respectively, assuming a spherical homogeneous dust grain.
The dust size, $a$, is fixed at 0.1~$\mu$m, and the absorption coefficients, $Q_{\rm abs, ol}$ and $Q_{\rm abs, pyr}$, are calculated from the optical constants of amorphous olivine ($\rm MgFeSiO_4$) and amorphous pyroxene ($\rm Mg_{0.5}Fe_{0.5}SiO_3$) obtained by the laboratory measurement of \citet{Dorschner1995} according to the Mie theory \citep{Bohren1998}. The mass densities of amorphous olivine and pyroxene, $\rho_{\rm ol}$ and $\rho_{\rm pyr}$, are 3.71 and 3.20~g~$\rm cm^{-3}$, respectively \citep{Dorschner1995}.
We use $\kappa_{\rm cry}$ of free-flying crystalline forsterite in aerosol measured by \citet{Tamanai2006}, the peak wavelengths of which are known to be consistent with the crystalline forsterite features observed in our Galaxy \citep{Wright2016}. The measurements using the aerosol technique are not quantitative, and hence we normalize $\kappa_{\rm cry}$ measured by \citet{Tamanai2006} assuming $\kappa_{\rm cry}(23~{\rm {\mu}m})=3.7{\times}10^3~{\rm cm^2g^{-1}}$ \citep{Fabian2001}.

The 11~$\mu$m crystalline feature is easily affected by the PAH emission feature at 11.3~$\mu$m, and hence it is hard to derive the crystallinity robustly by decomposing the 11~$\mu$m crystalline feature from the 10~$\mu$m amorphous feature. On the other hand, we clearly detect the 23~$\mu$m crystalline feature without being affected by the PAH emission, which is also pointed out in \citet{Stierwalt2014}.
Therefore we fix $N_{\rm cry}$ in the spectral fitting using the information on the 23~$\mu$m crystalline feature, which is determined by spectral fitting of the 23~$\mu$m feature with a Gaussian function (see the middle panel of Fig.~\ref{fig:fit_result}).
The absorption-free continuum for the 23~$\mu$m feature is assumed using a spline curve, the pivot points of which are anchored at 20, 21, 25.5 and 30~$\mu$m. The central wavelength is restricted to the wavelength range of 23.0--23.5~$\mu$m. 
In addition, as seen in the 16, 19 and 23~$\mu$m crystalline features in figure~5 of \citet{Spoon2006}, the crystalline features at longer wavelengths tend to show smaller apparent optical depths in the spectra of ULIRGs contrary to the laboratory measurements of $\kappa_{\rm cry}$, which is probably caused by the radiative transfer effects or the dilution by continuum emission from cold dust.
Hence we estimate $N_{\rm cry}$ for all the objects in our sample by calculating $CF_{\rm I08572}~{\times}~{\tau}_{\rm cry}({\rm 23~{\mu}m})/{\kappa}_{\rm cry}({\rm 11~{\mu}m})$, where $CF_{\rm I08572}$ is defined as $\tau_{\rm cry, I08572}(11~{\rm {\mu}m})/\tau_{\rm cry, I08572}(23~{\rm {\mu}m})$ and obtained from the spectral decomposition of the 10~$\mu$m silicate feature of IRAS~08572+3915 which shows little PAH emission.

It is difficult to obtain the uncertainty of the correction by $CF_{\rm I08572}$ because the crystalline features at $\sim$9.8 and $\sim$11.1~$\mu$m are diluted in almost all the sample spectra. On the other hand, the crystalline feature at $\sim$16~$\mu$m is not diluted much, and thus we investigate $\tau(16~{\rm {\mu}m})/\tau(23~{\rm {\mu}m})$, instead of $\tau(11~{\rm {\mu}m})/\tau(23~{\rm {\mu}m})$, for 12 sources shown in \citet{Spoon2006}. From figure~4 in \citet{Spoon2006}, the mean and standard deviation of the $\tau(16~{\rm {\mu}m})/\tau(23~{\rm {\mu}m})$ are found to be 1.71 and 0.46, respectively, and the value of $\tau(16~{\rm {\mu}m})/\tau(23~{\rm {\mu}m})$ for IRAS~08572+3915 is 1.46. 
We investigated the effect of the difference between 1.71 and 1.46 on the 5.3--12~$\mu$m spectral fitting, assuming $N_{\rm cry}$ 1.71/1.46 times larger. As a result, the results of the fits for all the sample did not change significantly with a significance level of 5\% by the F-test. Therefore we conclude that the uncertainty of the correction by $CF_{\rm I08572}$ does not affect the results of the 5.3--12~$\mu$m spectral fitting significantly. 

\subsection{Fitting procedure}
We performed the spectral fitting in the following four steps, in which we use the Levenberg–Marquardt algorithm \citep{levenberg, marquardt} for the $\chi^2$ minimization: (1) we determined the amplitudes of line emissions by spectral fitting for a narrow wavelength range assuming a quadratic function as a local continuum, and then fixed those parameters in the subsequent procedure.
(2) We tentatively fitted the spectrum of the wavelength range of 5.3--7.8~$\mu$m to obtain initial parameters for the following fitting processes. As for the fitting function, we assume equation~\ref{eq:model} setting $\tau_{\rm sil}$ and $F_{\rm hot, unobs}$ to be zero as the silicate feature is out of the wavelength range. Using the initial parameters thus obtained, we fitted the spectrum of the wavelength range of 5.3--12~$\mu$m.
(3) We obtained $CF_{\rm I08572}$ (defined in the previous subsection) by fitting the 5.3--12~$\mu$m spectrum of IRAS~08572+3915 with equation~\ref{eq:model} where free parameters are the mass column density of crystalline forsterite, $N_{\rm cry}$, as well as the parameters in Table~\ref{tab:para}. 
(4) We derived the properties of dust through the 5.3--12~$\mu$m spectral fitting to all the sample spectra. The free parameters are the same as in step (3) except that $N_{\rm cry}$ is fixed at $CF_{\rm I08572}~{\times}~{\tau}_{\rm cry}({\rm 23~{\mu}m})/{\kappa}_{\rm cry}({\rm 11~{\mu}m})$.

\begin{table*}
\caption{Free parameters in the 5.3--12~$\mu$m spectral fitting to all the sample spectra.}
\label{tab:para}     
\centering                      
\begin{tabular}{l c c}
\hline\hline                
 Component & Parameter & Description \\    
\hline 
$F_{\rm hot, AGN}$ & $A_{\rm hot, AGN}$ & Amplitude of the 5.3--8~$\mu$m hot dust continuum emission \\
& ${\Gamma}_{\rm hot, AGN}$ & Power-law index of the 5.3--8~$\mu$m hot dust continuum emission \\
$\tau_{\rm sil}$ & $N_{\rm ol}$ & Mass column density of amorphous olivine  \\
 & $N_{\rm pyr}$ & Mass column density of amorphous pyroxene  \\
$\tau_{\rm ice+HAC}$ & $\tau_{\rm ice, I00183}$ & Amplitude of the $\rm H_2O$ ice absorption template (IRAS~F00183--7111)  \\
 & $\tau_{\rm ice,~I12127}$ & Amplitude of the $\rm H_2O$ ice absorption template (IRAS~12127--1412)   \\
 & $\tau_{\rm ice,~N4418}$ & Amplitude of the $\rm H_2O$ ice absorption template (NGC~4418)   \\
 & $\tau_{\rm HAC,~6.85~{\mu}m}$ & Amplitude of the HAC absorption at 6.85~$\mu$m  \\
 & $\tau_{\rm HAC,~7.25~{\mu}m}$ & Amplitude of the HAC absorption at 7.25~$\mu$m  \\
$F_{\rm PAH}$ & $A_{\rm PAH}$ & Amplitude of the PAH emission  \\
 & $x_{\rm ion}$ & Ionization fraction of PAHs  \\
 & $A_{\rm PAH, 10.68}$ & Amplitude of the 10.68~$\mu$m PAH emission \\
 & $\tau_{\rm PAH}$ & Amplitude of the optical depth of the PAH emission \\
$F_{\rm hot, unobs}$ & $A_{\rm hot, unobs}$ & Amplitude of the unobscured hot dust continuum  \\

\hline 
\end{tabular}
\end{table*}

The right panel of Fig.~\ref{fig:fit_result} shows examples of the results of the spectral decomposition (see Appendix~\ref{sect:summary} for all the sample galaxies). 
  Although 20\% of the fits to the spectra, which have high signal-to-noise ratio, are not acceptable on the basis of the $\chi^2$ statistics with a significance level of 0.05, which is possibly caused mainly by neglecting the uncertainties of the absorption efficiency and the PAH model, the model is overall fitted to the spectra in our sample well enough to characterize the mineralogical composition and crystallinity. 
   The uncertainties of the fitting parameters are obtained by the Monte Carlo method of fitting 100 spectra with Gaussian noises.
   
 We consider whether a good fit can be achieved with significantly different sets of parameters in terms of parameter degeneracies. The spectral profile of the silicate absorption feature is determined by the mass ratio of amorphous olivine to pyroxene, $f_{\rm ol/pyr}=N_{\rm ol}/N_{\rm pyr}$, and the ratio of the unobscured to absorption-free continua, $f_{\rm unobs}=A_{\rm hot, unobs}/A_{\rm hot, AGN}$. We investigate the $\Delta \chi^2$ distribution within the parameter ranges of $f_{\rm ol/pyr} = [0, \infty]$ and $f_{\rm unobs}= [0, 1]$. 
In this analysis, free parameters are the total mass column density of silicate dust and the normalization of the continuum emission, while we fix the parameters of the power-law index of the continuum emission, the mass column density of crystalline forsterite, the amplitudes of the ice and HAC absorption templates, all the parameters associated with the PAH emission since they are determined almost uniquely based on the information outside the wavelength range of 8--12~$\mu$m and on the narrow PAH emission feature at 11.3~$\mu$m. From the $\Delta \chi^2$ distribution thus obtained, we confirm that the different sets of the parameters do not produce a significant fit in the spectral fitting for all the sample spectra.

\section{Results\label{sec:result}}

\begin{figure*}
	\centering
	\includegraphics[width=18cm]{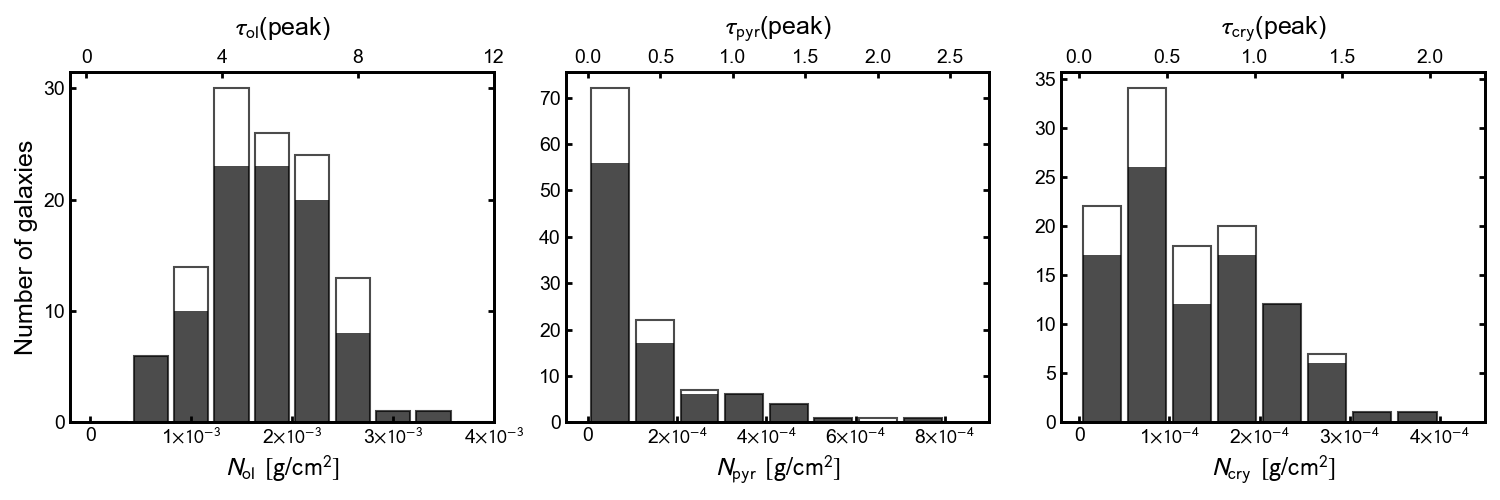}
	\caption{Histograms of the mass column densities of amorphous olivine ($N_{\rm ol}$; $\it left$), amorphous pyroxene ($N_{\rm pyr}$; $\it center$) and crystalline forsterite ($N_{\rm cry}$; $\it right$). The optical depths at the peak wavelengths corresponding to the mass column densities are shown in the upper x-axes. The peak wavelengths used for calculating the optical depths of amorphous olivine, amorphous pyroxene and crystalline forsterite are 9.8, 9.2 and 11.0~$\mu$m, respectively. Black and white bars indicate the sources whose 5.3--12~$\mu$m spectral fits are accepted and those rejected, respectively, on the basis of the $\chi^2$ statistics with a significance level of 0.05.}
	\label{fig:histo1}%
   \end{figure*}

 In Fig.~\ref{fig:histo1}, we show the histograms of the mass column densities of amorphous olivine, $N_{\rm ol}$, amorphous pyroxene, $N_{\rm pyr}$, and crystalline forsterite, $N_{\rm cry}$, obtained by the spectral fitting. We summarize the median values and the 25th and 75th percentiles in Table~\ref{tab:mean_n} (see Appendix~\ref{sect:summary} for all the sample galaxies).
 The histogram of $N_{\rm ol}$ in our sample shows a relatively symmetric distribution centered around $1.7{\times}10^{-3}~{\rm g~cm^{-2}}$, which corresponds to the peak optical depth of 5.5 for the 10~$\mu$m feature. On the other hand, the histograms of $N_{\rm pyr}$ and $N_{\rm cry}$ show asymmetric distributions with tails at larger column densities.
 Furthermore, we find that the sample galaxies have $N_{\rm ol}$ more than one order of magnitude higher than $N_{\rm pyr}$ and $N_{\rm cry}$ on average, and hence amorphous olivine is likely to be the main component of the silicate dust obscuring AGNs.
 
\begin{figure*}
	\centering
	\includegraphics[width=15cm]{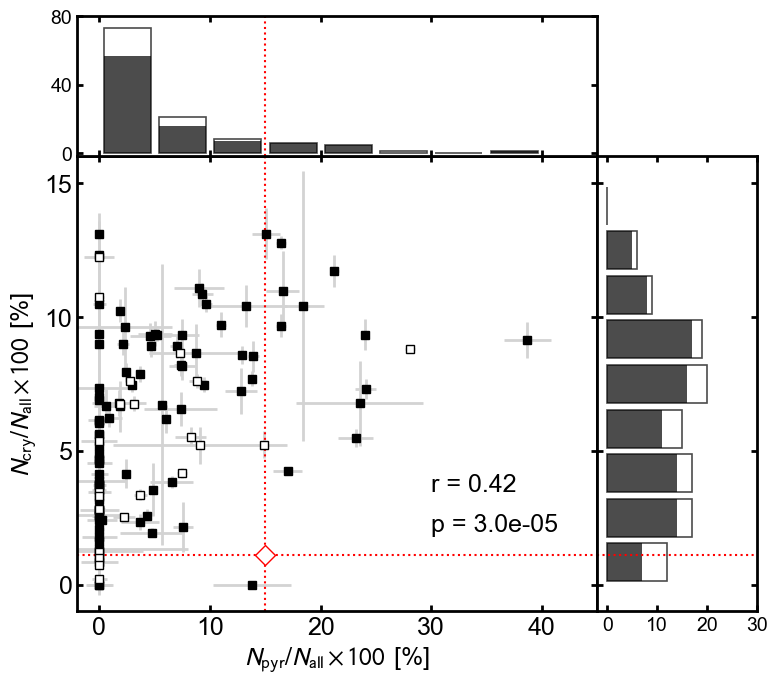}
	\caption{Correlation plot between $N_{\rm cry}/N_{\rm all}$ and $N_{\rm pyr}/N_{\rm all}$ and their histograms. Filled and open black squares show the objects whose 5.3--12~$\mu$m spectral fits are accepted and those rejected on the basis of the $\chi^2$ statistics with a significance level of 0.05, respectively. Black and white bars in the histograms also indicate the sources in which the spectral fits are accepted and rejected, respectively. The correlation coefficient and the p-value for the accepted sample are shown at the bottom right corner of the correlation plot. Red diamond and dotted lines show the silicate dust properties obtained from the Sgr~A* spectrum, a typical spectrum of the diffuse ISM in our Galaxy \citep{Kemper2004}.}
	\label{fig:histo}%
   \end{figure*}

We derive the abundances of the minor components by calculating the mass fractions, $N_{\rm pyr}/N_{\rm all}$ and $N_{\rm cry}/N_{\rm all}$, where $N_{\rm all}=N_{\rm ol}+N_{\rm pyr}+N_{\rm cry}$. In Fig.~\ref{fig:histo}, we show the histograms of the resultant $N_{\rm pyr}/N_{\rm all}$ and $N_{\rm cry}/N_{\rm all}$, the median fractions and the 25th and 75th percentiles of which are also summarized in Table~\ref{tab:mean_n}. 
We find that more than half of our sample have $N_{\rm pyr}/N_{\rm all}<5\%$, while $10\%$ of our sample have relatively high values $N_{\rm pyr}/N_{\rm all}>15\%$. The histogram of $N_{\rm cry}/N_{\rm all}$ shows a relatively uniform distribution from 0 to 13$\%$. 
Figure~\ref{fig:histo} also shows the presence of significant relationship between $N_{\rm cry}/N_{\rm all}$ and $N_{\rm pyr}/N_{\rm all}$, which indicates that the mass fractions of the minor components positively correlate to each other. 
On the other hand, $N_{\rm pyr}/N_{\rm all}$ and $N_{\rm cry}/N_{\rm all}$ do not correlate to the equivalent width of the PAH feature at 6.2~$\mu$m and the total optical depth of silicates with a significance level of 0.01.

\begin{table*}
\caption{General properties of the silicate dust in our sample AGNs}    
\label{tab:mean_n}     
\centering                      
\begin{tabular}{l c c c c c c c c}       
\hline\hline                
 & $N_{\rm ol}$~[${\rm g/cm^{2}}$] & ${\tau}_{\rm ol}$(peak) & $N_{\rm pyr}$~[${\rm g/cm^{2}}$] & ${\tau}_{\rm pyr}$(peak) & $N_{\rm cry}$~[${\rm g/cm^{2}}$] & ${\tau}_{\rm cry}$(peak) & $N_{\rm pyr}/N_{\rm all}$~[$\%$] & $N_{\rm cry}/N_{\rm all}$~[$\%$] \\    
\hline 
$\rm Median^{a}$ & $1.7{\times}10^{-3}$ & 5.5 & $5.0{\times}10^{-5}$ & 0.20 & $1.0{\times}10^{-4}$ & 0.38 &  2.2 & 6.2 \\
$\rm 25th^{b}$ & $1.4{\times}10^{-3}$ & 4.2 & 0 & 0 & $5.9{\times}10^{-5}$ & 0.22 & 0 & 3.3 \\
$\rm 75th^{c}$ & $2.2{\times}10^{-3}$ & 6.9 & $1.6{\times}10^{-4}$ & 0.67 & $1.8{\times}10^{-4}$ & 0.66 & 7.9 & 8.9 \\
\hline 
\end{tabular}
\tablefoot{\tablefoottext{a}{The median values.}
\tablefoottext{b}{The 25th percentiles.}
\tablefoottext{c}{The 75th percentiles.}
}
\end{table*}

\section{Discussion} 

\subsection{Properties of silicate dust in heavily obscured AGNs compared with those in our Galaxy \label{sec:overall}} 

The mineralogical composition in our Galaxy is observationally known to be variable depending on the environments where silicate dust exists. 
\citet{Demyk2001} suggest the evolution of amorphous silicate from amorphous olivine newly formed around evolved stars to amorphous pyroxene around young stars. 
Indeed, the 10~$\mu$m features of newly formed dust observed around evolved stars are fitted well with amorphous silicate with olivine-type stoichiometry \citep[e.g.,][]{Demyk2000, DoDuy2020}, while a typical spectrum of the diffuse ISM requires a small mass fraction of amorphous pyroxene \citep[olivine:pyroxene = 85:15 for the Sgr~A* spectrum;][]{Kemper2004}. 
Furthermore, it is suggested that the pyroxene mass fractions in molecular clouds and circumstellar regions of young stellar objects (YSOs) are higher than those of the diffuse ISM dust \citep[e.g.,][]{Demyk1999, vanBreemen2011, DoDuy2020}. 
Amorphous pyroxene is reported to be transformed from crystalline forsterite by ion irradiation \citep{Demyk2001, Rietmeijer2009}, and thus the ion bombardments can explain the evolution of amorphous silicate.
Pyroxene-type equilibrated dust is also formed from the ISM dust by vaporization and recondensation in dense and hot environments of circumsteller disks of YSOs \citep{Gail2004}. 
We show the pyroxene mass fraction of silicate dust toward Sgr~A* as a red vertical dotted line in the histogram of Fig.~\ref{fig:histo}, indicating an overall lower mass fraction of the amorphous pyroxene in the heavily obscured AGNs than that in the typical diffuse ISM in our Galaxy.
Accordingly, the silicate dust obscuring AGNs is likely to consist of newly formed dust originated from circumnuclear starburst activities.

In general, the crystallinity is known to be high in winds from evolved stars \citep[10-15$\%$; e.g.,][]{Kemper2001, Molster2002} and circumstellar regions surrounding YSOs \citep[$\sim15\%$; e.g.,][]{vanBoekel2005,Sargent2009}. 
On the other hand, crystalline silicate dust is almost absent in the diffuse ISM and molecular clouds \citep[$\sim$1-2$\%$; e.g.,][]{Kemper2004, vanBreemen2011, DoDuy2020}, and hence it is suggested that amorphization by the ion bombardments is important \citep[e.g.,][]{Demyk2001,Kemper2004}.
The high crystallinity of the silicate dust surrounding YSOs is considered to be caused by thermal annealing in the hot environment close to the central star \citep[e.g.,][]{vanBoekel2005,Sargent2009}. 
The average crystallinity of $\sim6\%$ in the heavily obscured AGNs in Table~\ref{tab:mean_n} and Fig.~\ref{fig:histo} is closer to those of newly formed dust around the evolved stars rather than the diffuse ISM dust in our Galaxy. It is possible that the overall high crystallinity of the silicate dust obscuring AGNs originates from evolved massive stars in the starburst environments, as also mentioned in \citet{Spoon2006}.
The crystallinity is determined by the balance of dust production, amorphization and destruction rates.
\citet{Kemper2011} simulate whether the starburst activities alone can explain the crystallinities of 6.5--13\% suggested in \citet{Spoon2006}, and show that such a high crystallinity cannot be explained unless extreme input parameters are assumed for such as the star formation rate and the initial crystallinity in the stellar ejecta. If a top heavy initial mass function (IMF) \citep{Zhang2018} is assumed instead of the IMF in e.g., \citet{Kroupa2001}, high crystallinity may be achieved temporarily because the starburst activity produces large amounts of dust on a shorter timescale than that for amorphization. We do not consider this possibility in the following discussion.
Accordingly, in order to explain the overall trend of high crystallinity, we consider re-crystallization of amorphous silicate or production of crystalline silicate other than starburst activities.

Although crystallization occurs in high-temperature environments at $\sim$1000~K, the crystalline silicate in heavily obscured AGNs is considered to be located in relatively outer cooler regions because we detect the crystalline features only in the absorption at the wavelength of 23~$\mu$m. 
For a re-crystallization mechanism of amorphous silicate in cooler regions in heavily obscured AGNs, we consider a possibility of a localized crystallization due to a transient heating by shock waves \citep{Harker2002}, driven by supernovae in the circumnuclear starburst and/or outflow if the shocks can anneal silicate dust up to $\sim$1000~K. Because the chemical equilibrium cannot be achieved by a short annealing time typical of the shock heating, crystalline enstatite, the Mg end member of pyroxene, is unlikely to be formed \citep{Gail2004}, which is consistent with the spectral characteristics of our sample spectra. 
Ions accelerated in shock wave propagation and cosmic rays, which are considered to be accelerated in the shock front \citep[e.g.,][]{Bell1978}, can destroy the crystalline structure of silicate dust \citep{Demyk2001}, which negatively acts for the enrichment of the crystalline silicate. 
Therefore we consider that the localized crystallization due to a transient heating by shock waves is hard to cause the high crystallinity.

Another possibility is that the crystalline silicate is produced or processed in the high temperature environments close to the nucleus and transported to outer cooler region by radial mixing or outflow.
Indeed, \citet{Kemper2007} report the emission feature due to crystalline forsterite in a quasar wind spectrum.
However, \citet{Spoon2006} conclude that this possibility is unlikely because of the following two reasons: one is that crystalline feature is not detected in the central 2~pc of NGC~1068 in the mid-IR interferometric observation in \citet{Jaffe2004}. \citet{Raban2009} report the follow-up observation of NGC~1068 with a different position angle and longer baseline length than those in \citet{Jaffe2004}. We find that the 11~$\mu$m crystalline feature is significantly detected in the mid-IR spectra shown in \citet{Raban2009} although \citet{Raban2009} do not mention the crystalline feature. Hence the crystallinity of the central region of NGC1068 rather supports the picture that crystalline silicate originates from AGNs.
The other reason is that we have to newly introduce a large-scale transportation mechanism itself.
Recent dynamical views of an AGN torus \citep[e.g.,][]{Wada2012} are likely to explain a large-scale transportation. Indeed, molecular outflows have been observed in nearby obscured AGNs including a lot of objects in our sample in the recent years \citep[e.g.,][]{Aalto2012, Veilleux2013, Lutz2020}. For example, NGC~1377 in our sample is known to have a 150~pc-scale molecular outflow. The large mass outflow rate suggests the presence of a feedback loop of cyclic outflow \citep[e.g.,][]{Aalto2020}.
The projected size of the 800~$\mu$m nuclear dust continuum of NGC~1377 is observed to be ${\sim}4$~pc \citep{Aalto2020}. The timescale for amorphization due to the cosmic-ray bombardment is estimated to be 70~Myr \citep{Bringa2007}. Thus the dust radial velocity needs to be faster than 0.05~km/s, which is considerably slower as compared to the outflow velocity of the molecular wind, 90~km/s, measured by \citet{Aalto2020}. \citet{Kozasa1999} report that it takes only 30 days processing into crystalline forsterite at the dust temperature of 1000 K, which is negligibly short. 
Thus it is possible that crystalline silicate processed around the nucleus is transported to outer regions by the feedback loop of outflow.

\subsection{Origin of the difference in the silicate dust properties}

In Fig.~\ref{fig:histo}, we find the positive correlation between $N_{\rm cry}/N_{\rm all}$ and $N_{\rm pyr}/N_{\rm all}$, indicating that heavily obscured AGNs with a relatively higher mass fraction of the amorphous pyroxene, which deviate from the overall trend, show a higher mass fraction of the crystalline forsterite.
According to the above discussion, the amorphous olivine, the main component of silicate obscuring AGNs, is crystallized by the thermal annealing close to the nucleus and transported to the outer region.
The mass fraction of the crystalline forsterite is likely to vary depending on the degree of the thermal processing or the radial transportation of silicate dust among the heavily obscured AGNs. Therefore heavily obscured AGNs with high mass fractions of the crystalline forsterite is naturally of high mass fractions of amorphous pyroxene considering that crystalline forsterite is partially transformed to amorphous pyroxene by ion bombardments in/after the radial transportation to the outer cooler region as the main formation mechanism of amorphous pyroxene. For another formation mechanism, we consider that crystalline enstatite, which is likely to be formed from the ISM dust by vaporization and recondensation \citep{Gail2004} in the dense and hot nuclear environment, is processed to amorphous pyroxene by cosmic-ray bombardments. However, the spectral features of crystalline enstatite are not detected in the spectra of our sample, and therefore we excluded this possibility.
We also find that 5 out of 15 sources with a mass fraction of >10\% crystalline forsterite, which do not have a significant fraction of amorphous pyroxene, deviate from the trend of the positive correlation. Since the timescale for amorphization is considerably longer than that for the dust radial transportation as discussed in Sect.~\ref{sec:overall}, crystalline forsterite is likely to amorphize mainly after the transportation. Hence, if we observe the sources just after the radial transportation started, the mass fraction of crystalline forsterite is likely to be high without producing a significant fraction of amorphous pyroxene. 
Accordingly, the difference in the silicate dust properties among the heavily obscured AGNs originates mainly from the difference in the degree of the AGN activities, that is, the thermal processing or the radial transportation.

\begin{figure*}
	\centering
	\includegraphics[width=13cm]{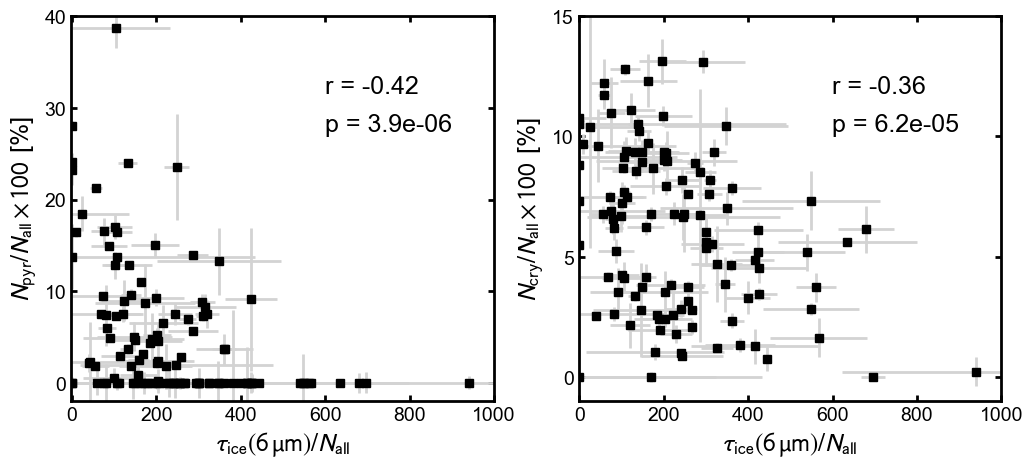}
	\caption{Anti-correlations of $\tau_{\rm ice}(6~{\rm {\mu}m})/N_{\rm all}$ with $N_{\rm cry}/N_{\rm all}$ ({\it left}) and $N_{\rm pyr}/N_{\rm all}$ ({\it right}). The correlation coefficients and the p-values for the sample are shown at the top right corners. }
	\label{fig:ice}%
   \end{figure*}

We focus on the temperature of the circumnuclear region to confirm that the difference in $N_{\rm cry}/N_{\rm all}$ and $N_{\rm pyr}/N_{\rm all}$ originates mainly from the difference in the degree of the thermal processing.
In general, $\rm H_2O$ ice, the sublimation temperature of which is 90~K \citep{Tielens2005}, is assumed to exist in the mantle of dust in dense cold molecular clouds.
Hence, for heavily obscured AGNs with deep ice absorptions at 3.0 and 6.0~$\mu$m, silicate dust in dense cold circumnuclear regions is likely to dominate the total silicate absorption along the line of sight. 
Icy dust is expected to exist in the relatively warm environment since the $\rm CO_2$ ice absorption feature at 15.0~$\mu$m, the sublimation temperature of which is 50~K \citep{Tielens2005}, are not detected for most of the ULIRGs by \citet{Lahuis2007} in contrast to the detection of the 6~$\mu$m $\rm H_2O$ ice feature. In our Galaxy, deep $\rm CO_2$ ice absorption features are observed in dense molecular clouds \citep{Gibb2000}, and thus we conclude that dense molecular clouds in the host galaxies do not contribute to the total column density of $\rm H_2O$ ice.
We calculate the ratio of the 6~$\mu$m optical depth of $\rm H_2O$ ice to the total mass column density of silicate dust, $\tau_{\rm ice}(6~{\rm {\mu}m})/N_{\rm all}$, which is likely to depend on the average temperature of obscuring dust.
In Fig.~\ref{fig:ice}, we compare the silicate properties shown in Fig.~\ref{fig:histo} with $\tau_{\rm ice}(6~{\rm {\mu}m})/N_{\rm all}$, which indicates that the obscuring silicate dust in galaxies with high $\tau_{\rm ice}(6~{\rm {\mu}m})/N_{\rm all}$ tends to be of low mass fractions of amorphous pyroxene and crystalline forsterite. 
Hence silicate dust obscuring AGNs with a high $\tau_{\rm ice}(6~{\rm {\mu}m})/N_{\rm all}$, which has a relatively low temperature, is considered to be processed less heavily.
Assuming the nucleus as the heating source, cooler dust is located more distant from the nucleus, and therefore is harder to be transported from the nucleus. 
This is consistent with the above discussion, and thus dust processing is likely to originate from the AGN activity.  

\section{Conclusions}
We investigate AGN activities obscured by large amounts of dust through a systematic study of silicate dust properties.
We select 115 mid-IR spectra of the heavily obscured AGNs observed by Spitzer/IRS, and characterize the silicate features by spectral fitting focusing on the mineralogical composition and crystallinity.
As a result, we fit the spectra well using three dust components of amorphous olivine, amorphous pyroxene and crystalline forsterite.
We find that the main component of silicate dust obscuring AGNs is amorphous olivine, the average mass column density of which is one order of magnitude higher than those of amorphous pyroxene and crystalline forsterite. We also find that the mass fractions of amorphous pyroxene and crystalline forsterite significantly correlate to each other.

The overall dust properties of heavily obscured AGNs tend to be of a lower mass fraction of the amorphous pyroxene and of a higher mass fraction of the crystalline forsterite than those of the diffuse ISM in our Galaxy.
The overall trend of the low mass fraction of the amorphous pyroxene suggests that silicate dust newly formed in starburst activities is abundant for the obscuring silicate dust. 
The overall trend of high crystallinity and the fact that crystalline features are seen only in the absorption suggest that amorphous silicate is thermally processed close to the nucleus and radially transported to an outer cooler region.
Considering that amorphous pyroxene can be transformed from crystalline forsterite by ion bombardments, we can explain the positive correlation between the amorphous pyroxene and crystalline forsterite mass fractions. 
Therefore the difference in the dust properties among the heavily obscured AGNs originates mainly from the difference in the degree of the thermal processing or the radial transportation.
We also find that heavily obscured AGNs with high ratios of the $\rm H_2O$ ice absorption to silicate mass column density tend to have silicate dust with lower mass fractions of amorphous pyroxene and crystalline forsterite, which is consistent with the scenario of the thermal dust processing close to the nucleus.

\begin{acknowledgements}
This work is based on observations with the Spitzer Space Telescope, which is operated by the Jet Propulsion Laboratory, California Institute of Technology under a contract with NASA, using the the Combined Atlas of Sources with Spitzer IRS Spectra (CASSIS). CASSIS is a product of the IRS instrument team, supported by NASA and JPL.  
\end{acknowledgements}

%
   \bibliographystyle{aa} 
   \bibliography{IRS_silicate} 
%
\appendix

\section{Sharpness of the 10~$\mu$m silicate feature\label{sect:sharp}}

\begin{figure}
	\centering
	\includegraphics[width=9cm]{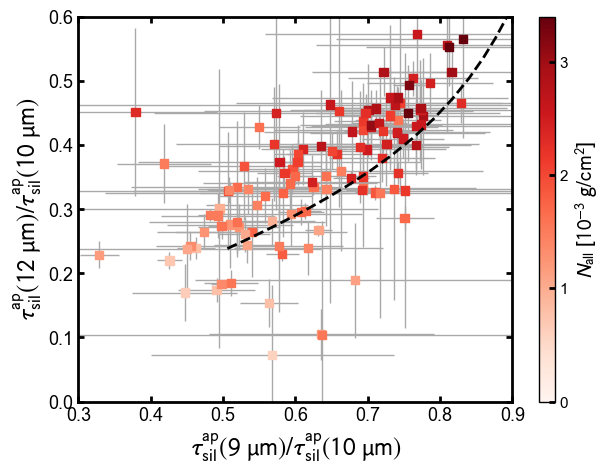}
	\caption{Relation between $\tau_{\rm sil}^{\rm ap}(12~{\rm {\mu}m})/\tau_{\rm sil}^{\rm ap}(10~{\rm {\mu}m})$ and $\tau_{\rm sil}^{\rm ap}(9~{\rm {\mu}m})/\tau_{\rm sil}^{\rm ap}(10~{\rm {\mu}m})$ color-coded by the total mass column density of silicate dust, $N_{\rm all}$. Dashed line shows the resultant optical depth ratios obtained by the  simulation of the saturation effects due to the unobscured hot dust emission.}
	\label{fig:tau_ratio}%
   \end{figure}

We consider a possibility that dust properties explain the difference in the sharpness of the 10~$\mu$m silicate feature instead of that of the saturation due to the unobscured emission component.
Broad 10~$\mu$m emission features seen in the spectra of type-1 AGNs are explained by larger-size or higher-porosity dust than typical 0.1~$\mu$m-sized homogeneous dust \citep[e.g., ][]{Xie2017, Li2008}, while those seen in the spectra of circumstellar disks are explained by, for instance, the combination of silica ($\rm SiO_2$) at 8.6~$\mu$m, crystalline enstatite ($\rm MgSiO_3$), forsterite and amorphous features \citep[e.g.,][]{vanBoekel2005, Olofsson2009}. 
In order to identify the causes of the difference in the sharpness, we calculate the two apparent optical depth ratios $\tau_{\rm sil}^{\rm ap}(9~{\rm {\mu}m})/\tau_{\rm sil}^{\rm ap}(10~{\rm {\mu}m})$ and $\tau_{\rm sil}^{\rm ap}(12~{\rm {\mu}m})/\tau_{\rm sil}^{\rm ap}(10~{\rm {\mu}m})$ using the optical depths $\tau_{\rm sil}^{\rm ap}(\lambda)$ determined by the absorption-free continuum of $F_{\rm hot, AGN}$. Both smaller $\tau_{\rm sil}^{\rm ap}(9~{\rm {\mu}m})/\tau_{\rm sil}^{\rm ap}(10~{\rm {\mu}m})$ and $\tau_{\rm sil}^{\rm ap}(12~{\rm {\mu}m})/\tau_{\rm sil}^{\rm ap}(10~{\rm {\mu}m})$ indicate sharper 10~$\mu$m feature if the central wavelength of the feature is at around 10~$\mu$m. Figure~\ref{fig:tau_ratio} shows the relationship between $\tau_{\rm sil}^{\rm ap}(12~{\rm {\mu}m})/\tau_{\rm sil}^{\rm ap}(10~{\rm {\mu}m})$ and $\tau_{\rm sil}^{\rm ap}(9~{\rm {\mu}m})/\tau_{\rm sil}^{\rm ap}(10~{\rm {\mu}m})$, which clearly shows a tight positive correlation. If the size or porosity of dust caused the difference in the sharpness, a negative correlation between the ratios should be observed because the central wavelength of the 10~$\mu$m feature shifts towards longer wavelengths due to larger-sized or higher-porosity amorphous silicate. Therefore it is unlikely that the difference in the sharpness is caused by the size or porosity of dust.

We simulate the saturation effects due to the unobscured hot dust emission using the absorption efficiency of amorphous olivine, $Q_{\rm abs, ol}$, which is the main component of the silicate dust in our sample, by changing the contribution levels of the unobscured component. We assume the mass column density of the silicate dust of $1.7{\times}10^{-3}~{\rm g/cm^2}$, the mean mass column density of amorphous olivine in our sample obtained by the spectral fitting, for the simulation.
The relationship between $\tau_{\rm sil}^{\rm ap}(12~{\rm {\mu}m})/\tau_{\rm sil}^{\rm ap}(10~{\rm {\mu}m})$ and $\tau_{\rm sil}^{\rm ap}(9~{\rm {\mu}m})/\tau_{\rm sil}^{\rm ap}(10~{\rm {\mu}m})$ (the dashed line in Fig.~\ref{fig:tau_ratio}) traces the trend of the data points well, although there is a slight offset presumably due to the presence of PAH or line features. 
Figure~\ref{fig:tau_ratio} is color-coded according to the mass column density of silicate dust, $N_{\rm all}$ obtained by the 5.3--12~$\mu$m spectral fitting. We find that the spectra with deeper absorptions show flatter profiles at the bottom of the feature. 
When the saturation causes the difference in the sharpness, the trend is reasonable considering that the ratio of the unobscured hot dust emission to absorption-free continuum emission does not change much.
Therefore we conclude that the unobscured emission component is likely to cause the difference in the sharpness.

In the 5.3--12~$\mu$m spectral fitting, we assume the spectral profile of the absorption-free continuum emission  due  to  the  hot  dust  heated  by  AGN, not the extended star-formation component responsible for the PAH emission. We verify the validity of this assumption. We investigate the relationship between the equivalent width of the PAH feature at 6.2~$\mu$m and the ratio of $A_{\rm hot, unobs}$ to $A_{\rm hot, AGN}$, which is expected to correlate positively if the weak continuum filling in the 9.7~$\mu$m feature is attributed to the extended star-formation component. The correlation coefficient and the p-value are 0.08 and 0.37, respectively, which do not indicate a significant correlation. Moreover, we also compare the results of the spectral fittings using the unobscured hot dust continua due to the starburst and AGN activities. We use the spectral profile of the dust continuum model obtained from NGC~7714 by \citet{Marshall2018} for the former continuum profile. As a result, for 91/115 sources, the latter model is fitted better than the former, and, for 8/115 sources, the reduced $\chi^2$ value improves significantly with a significance level of 5\% by the F-test. Hence we conclude that the unobscured AGN component fills the 9.7~$\mu$m feature.

\section{Summary of the analyses of the silicate features\label{sect:summary}}
The properties of silicate dust in our sample are summarized in Table~\ref{tab:summary}.
Figure~\ref{fig:app_mineral1} shows all the results of the spectral decomposition.

\begin{table*}[h]
\caption{Summary of the spectral fittings.}             
\label{table:result}     
\centering     
 \begin{tabular}{l c c c c c}        
\hline\hline                 
 Name & $N_{\rm ol}$~[$10^{-4}~{\rm g/cm^{2}}$] & $N_{\rm pyr}$~[$10^{-5}~{\rm g/cm^{2}}$] & $N_{\rm cry}$~[$10^{-6}~{\rm g/cm^{2}}$] & $N_{\rm pyr}/N_{\rm all}$~[$\%$] & $N_{\rm cry}/N_{\rm all}$~[$\%$] \\    
 (1)&(2)&(3)&(4)&(5)&(6)\\
\hline

IRAS~00091-0738 & 23.4~$\pm$~0.6 & 14~$\pm$~2 & 25.63~$\pm$~0.08 &   5.2~$\pm$~0.7 &   9.3~$\pm$~0.2 \\
IRAS~F00183-7111 & 18.2~$\pm$~0.2 & 31.9~$\pm$~0.8 & 17.8~$\pm$~0.6 &  13.8~$\pm$~0.4 &   7.7~$\pm$~0.3 \\
IRAS~00188-0856 & 10~$\pm$~2 &  0.0~$\pm$~0.7 &  0.2~$\pm$~0.6 &   0.0~$\pm$~0.7 &   0.2~$\pm$~0.6 \\
IRAS~00397-1312 & 19.7~$\pm$~0.5 &  4.0~$\pm$~0.1 & 14.6~$\pm$~0.7 &  1.85~$\pm$~0.06 &   6.8~$\pm$~0.4 \\
IRAS~00406-3127 & 15~$\pm$~1 &  0~$\pm$~1 &  6~$\pm$~1 &   0.0~$\pm$~0.6 &   3.5~$\pm$~0.9 \\
IRAS~01166-0844SE & 22.7~$\pm$~0.6 & 49~$\pm$~4 & 12.3~$\pm$~0.1 & 17~$\pm$~1 &   4.3~$\pm$~0.1 \\
IRAS~F01173+1405 & 14.4~$\pm$~0.1 &  5.0~$\pm$~0.6 & 10.8~$\pm$~0.5 &   3.2~$\pm$~0.4 &   6.8~$\pm$~0.3 \\
IRAS~F01197+0044 & 10.5~$\pm$~0.4 &  0.0~$\pm$~0.5 & 13~$\pm$~1 &   0.0~$\pm$~0.4 &  10.8~$\pm$~0.9 \\
IRAS~01199-2307 & 19.9~$\pm$~0.4 & 32.7~$\pm$~0.6 & 21.8~$\pm$~0.8 &  12.9~$\pm$~0.3 &   8.6~$\pm$~0.3 \\
IRAS~01298-0744 & 27~$\pm$~1 & 14~$\pm$~6 & 28.7~$\pm$~0.8 &  5~$\pm$~2 &   9.3~$\pm$~0.5 \\
IRAS~01355-1814 & 12.8~$\pm$~0.2 & 40.4~$\pm$~0.5 & 22~$\pm$~1 &  21.2~$\pm$~0.4 &  11.7~$\pm$~0.6 \\
IRAS~F01478+1254 & 19~$\pm$~3 &  5~$\pm$~9 & 21~$\pm$~2 &  2~$\pm$~4 & 10~$\pm$~2 \\
IRAS~01569-2939 & 20.6~$\pm$~0.4 & 25~$\pm$~1 & 27.0~$\pm$~0.6 &   9.6~$\pm$~0.6 &  10.5~$\pm$~0.3 \\
IRAS~02438+2122 & 13.6~$\pm$~0.1 &  0.0~$\pm$~0.6 &  3.9~$\pm$~0.2 &   0.0~$\pm$~0.4 &   2.8~$\pm$~0.1 \\
IRAS~02455-2220 & 13.95~$\pm$~0.07 & 0 &  8~$\pm$~1 & 0 &  6~$\pm$~1 \\
IRAS~02530+0211 & 12.2~$\pm$~0.4 & 40~$\pm$~2 &  9.4~$\pm$~0.5 & 23~$\pm$~2 &   5.5~$\pm$~0.3 \\
IRAS~03158+4227 & 22~$\pm$~1 &  0~$\pm$~4 &  1.9~$\pm$~0.4 &  0~$\pm$~2 &   0.9~$\pm$~0.2 \\
NGC~1377 & 14.1~$\pm$~0.1 & 62.7~$\pm$~0.9 & 19.71~$\pm$~0.08 &  28.1~$\pm$~0.5 &  8.82~$\pm$~0.07 \\
IRAS~03538-6432 & 14~$\pm$~2 &  0~$\pm$~3 &  4~$\pm$~3 &  0~$\pm$~2 &  3~$\pm$~2 \\
IRAS~03582+6012 & 27.6~$\pm$~0.5 &  6~$\pm$~3 &  7.4~$\pm$~0.3 &  2~$\pm$~1 &   2.5~$\pm$~0.1 \\
IRAS~04074-2801 & 21.5~$\pm$~0.8 & 13~$\pm$~4 & 23.6~$\pm$~0.8 &  5~$\pm$~2 &   9.4~$\pm$~0.5 \\
IRAS~04313-1649 & 24.3~$\pm$~0.9 &  1~$\pm$~3 &  6.1~$\pm$~0.9 &  0~$\pm$~1 &   2.4~$\pm$~0.4 \\
IRAS~04384-4848 & 15.6~$\pm$~0.1 &  0.0~$\pm$~0.2 & 10~$\pm$~1 &   0.0~$\pm$~0.1 &   6.0~$\pm$~0.7 \\
ESO~203-IG001 & 19.6~$\pm$~0.4 & 46~$\pm$~1 & 35.5~$\pm$~0.4 &  16.4~$\pm$~0.5 &  12.8~$\pm$~0.2 \\
IRAS~05020-2941 & 22.9~$\pm$~0.8 & 27~$\pm$~3 & 31.1~$\pm$~0.8 &  9~$\pm$~1 &  10.8~$\pm$~0.4 \\
IRAS~F06076-2139 & 11.0~$\pm$~0.1 &  0.0~$\pm$~0.5 &  0.8~$\pm$~0.5 &   0.0~$\pm$~0.4 &   0.8~$\pm$~0.5 \\
IRAS~06206-6315 & 12~$\pm$~2 &  0~$\pm$~4 &  9.2~$\pm$~0.6 &  0~$\pm$~3 &  7~$\pm$~1 \\
IRAS~06301-7934 & 19.8~$\pm$~0.2 & 27~$\pm$~1 & 24~$\pm$~1 &  11.0~$\pm$~0.4 &   9.7~$\pm$~0.5 \\
IRAS~06361-6217 & 12.9~$\pm$~0.7 & 0 &  9.6~$\pm$~0.6 & 0 &   6.9~$\pm$~0.5 \\
IRAS~F07224+3003 & 17.4~$\pm$~0.3 &  7~$\pm$~1 &  6.3~$\pm$~0.4 &   3.7~$\pm$~0.6 &   3.4~$\pm$~0.2 \\
IRAS~07251-0248 & 24.2~$\pm$~0.6 & 0 &  3.0~$\pm$~0.5 & 0 &   1.2~$\pm$~0.2 \\
MCG~+02-20-003 & 23.9~$\pm$~0.2 &  0~$\pm$~1 &  2.4~$\pm$~0.4 &   0.0~$\pm$~0.4 &   1.0~$\pm$~0.2 \\
SDSS~J082001.72+505039.1 &  8~$\pm$~2 & 13~$\pm$~3 &  0.00~$\pm$~0.04 & 14~$\pm$~4 &  0.00~$\pm$~0.04 \\
IRAS~08201+2801 & 25~$\pm$~1 &  5~$\pm$~7 & 18.9~$\pm$~0.8 &  2~$\pm$~2 &   6.8~$\pm$~0.5 \\
IRAS~F08520-6850 & 17.0~$\pm$~0.3 & 19.3~$\pm$~0.3 & 15.3~$\pm$~0.5 &   9.5~$\pm$~0.2 &   7.5~$\pm$~0.3 \\
IRAS~08572+3915 & 13~$\pm$~1 &  0~$\pm$~2 & 18.7~$\pm$~0.4 &  0~$\pm$~1 & 12~$\pm$~1 \\
IRAS~09039+0503 & 23~$\pm$~1 & 16~$\pm$~5 &  9.7~$\pm$~0.2 &  7~$\pm$~2 &   3.8~$\pm$~0.2 \\
IRAS~09539+0857 & 24~$\pm$~1 &  5.1~$\pm$~0.6 & 28.0~$\pm$~0.4 &   1.9~$\pm$~0.2 &  10.2~$\pm$~0.4 \\
IRAS~F10038-3338 & 27.15~$\pm$~0.09 & 23.5~$\pm$~0.6 & 28.0~$\pm$~0.4 &   7.3~$\pm$~0.2 &   8.7~$\pm$~0.1 \\
IRAS~10091+4704 & 31.7~$\pm$~0.3 &  8.30~$\pm$~0.09 & 14~$\pm$~2 &  2.44~$\pm$~0.04 &   4.1~$\pm$~0.5 \\
IRAS~F10112-0040 & 11~$\pm$~1 &  0.0~$\pm$~0.9 &  4.6~$\pm$~0.5 &   0.0~$\pm$~0.8 &   4.1~$\pm$~0.6 \\
IRAS~10173+0828 & 23.4~$\pm$~0.5 & 0 &  9.1~$\pm$~0.1 & 0 &  3.75~$\pm$~0.09 \\
IRAS~F10237+4720 & 27.1~$\pm$~0.8 & 43~$\pm$~5 & 24~$\pm$~3 & 13~$\pm$~1 &   7.2~$\pm$~0.9 \\
IRAS~10378+1109 & 23.2~$\pm$~0.8 &  9~$\pm$~4 &  5.8~$\pm$~0.7 &  4~$\pm$~2 &   2.3~$\pm$~0.3 \\
IRAS~10485-1447 & 13.7~$\pm$~0.3 & 0 &  8.2~$\pm$~0.1 & 0 &   5.6~$\pm$~0.1 \\
IRAS~11028+3130 & 24~$\pm$~1 &  5.0~$\pm$~0.2 & 17~$\pm$~2 &   1.9~$\pm$~0.1 &  7~$\pm$~1 \\
IRAS~11038+3217 & 19.6~$\pm$~0.4 & 16~$\pm$~2 &  5~$\pm$~2 &   7.5~$\pm$~0.9 &   2.2~$\pm$~0.9 \\
IRAS~11095-0238 & 23.6~$\pm$~0.5 &  7.6~$\pm$~0.2 & 19.7~$\pm$~0.5 &  2.91~$\pm$~0.09 &   7.5~$\pm$~0.3 \\
IRAS~11130-2659 & 21~$\pm$~1 & 23~$\pm$~6 & 28.6~$\pm$~0.8 &  9~$\pm$~2 &  11.1~$\pm$~0.7 \\
IRAS~11180+1623 & 19.5~$\pm$~0.5 & 35.0~$\pm$~0.9 & 21~$\pm$~1 &  13.9~$\pm$~0.5 &   8.5~$\pm$~0.6 \\
IRAS~11223-1244 & 11.7~$\pm$~0.6 &  0.0~$\pm$~0.2 &  6.4~$\pm$~0.9 &   0.0~$\pm$~0.2 &   5.2~$\pm$~0.8 \\
IRAS~11506+1331 & 11.7~$\pm$~0.4 & 0 & 17.7~$\pm$~0.3 & 0 &  13.1~$\pm$~0.5 \\
IRAS~11524+1058 & 18~$\pm$~1 &  0~$\pm$~4 &  7~$\pm$~2 &  0~$\pm$~2 &  4~$\pm$~1 \\
IRAS~11582+3020 & 23.3~$\pm$~0.2 & 16.0~$\pm$~0.1 & 16~$\pm$~1 &  6.02~$\pm$~0.07 &   6.2~$\pm$~0.5 \\
IRAS~12032+1707 & 20.4~$\pm$~0.2 &  0~$\pm$~1 &  6.0~$\pm$~0.7 &   0.0~$\pm$~0.5 &   2.8~$\pm$~0.4 \\
\hline

\end{tabular}

 \label{tab:summary}
 \end{table*}

\setcounter{table}{0}
\begin{table*}[h]
\caption{Continued.}            
\centering     
 \begin{tabular}{l c c c c c}        
\hline\hline                 
 Name & $N_{\rm ol}$~[$10^{-4}~{\rm g/cm^{2}}$] & $N_{\rm pyr}$~[$10^{-5}~{\rm g/cm^{2}}$] & $N_{\rm cry}$~[$10^{-6}~{\rm g/cm^{2}}$] & $N_{\rm pyr}/N_{\rm all}$~[$\%$] & $N_{\rm cry}/N_{\rm all}$~[$\%$] \\    
 (1)&(2)&(3)&(4)&(5)&(6)\\
\hline

IRAS~12127-1412 &  7.71~$\pm$~0.07 & 0 &  4.0~$\pm$~0.2 & 0 &   4.9~$\pm$~0.2 \\
IRAS~F12224-0624 & 33.2~$\pm$~0.2 &  0~$\pm$~1 &  7~$\pm$~1 &   0.0~$\pm$~0.4 &   2.1~$\pm$~0.3 \\
NGC~4418 & 20.0~$\pm$~0.1 &  6.2~$\pm$~0.9 & 17.0~$\pm$~0.3 &   2.8~$\pm$~0.4 &   7.6~$\pm$~0.1 \\
IRAS~12359-0725 & 14.0~$\pm$~0.8 & 0 & 10.5~$\pm$~0.9 & 0 &   7.0~$\pm$~0.7 \\
IRAS~12447+3721 & 16.0~$\pm$~0.3 & 56~$\pm$~2 & 17.1~$\pm$~0.8 & 24~$\pm$~1 &   7.3~$\pm$~0.4 \\
IRAS~F13045+2354 &  9.2~$\pm$~0.4 & 0 & 13~$\pm$~1 & 0 & 12~$\pm$~1 \\
IRAS~13106-0922 & 27~$\pm$~2 &  0~$\pm$~11 &  3~$\pm$~2 &  0~$\pm$~4 &   1.3~$\pm$~0.7 \\
IRAS~F13279+3401 & 20.0~$\pm$~0.2 & 71.9~$\pm$~0.7 & 28~$\pm$~2 &  24.0~$\pm$~0.3 &   9.3~$\pm$~0.6 \\
IRAS~13352+6402 & 13.2~$\pm$~0.3 &  7.0~$\pm$~0.2 &  5~$\pm$~1 &   4.9~$\pm$~0.2 &  4~$\pm$~1 \\
Mrk~273 & 14.8~$\pm$~0.8 & 14~$\pm$~2 &  9.5~$\pm$~0.4 &  8~$\pm$~1 &   5.5~$\pm$~0.3 \\
IRAS~14070+0525 & 25.2~$\pm$~0.2 &  0~$\pm$~1 &  8~$\pm$~1 &   0.0~$\pm$~0.5 &   3.2~$\pm$~0.5 \\
IRAS~14121-0126 & 16.1~$\pm$~0.5 &  0.0~$\pm$~0.3 &  3~$\pm$~1 &   0.0~$\pm$~0.2 &   1.6~$\pm$~0.8 \\
IRAS~F14242+3258 & 13.6~$\pm$~0.3 &  8.8~$\pm$~0.2 & 10~$\pm$~8 &   5.7~$\pm$~0.3 &  7~$\pm$~5 \\
IRAS~14348-1447 & 21.0~$\pm$~0.5 &  0~$\pm$~2 &  7.5~$\pm$~0.5 &  0~$\pm$~1 &   3.5~$\pm$~0.2 \\
IRAS~F14394+5332 & 16~$\pm$~1 &  0~$\pm$~3 &  4.2~$\pm$~0.4 &  0~$\pm$~2 &   2.6~$\pm$~0.3 \\
IRAS~F14511+1406 &  9.4~$\pm$~0.2 &  0.0~$\pm$~0.7 &  2~$\pm$~1 &   0.0~$\pm$~0.7 &  3~$\pm$~1 \\
IRAS~F14554+3858 & 13.2~$\pm$~0.9 &  0~$\pm$~2 &  0.0~$\pm$~0.3 &  0~$\pm$~1 &   0.0~$\pm$~0.2 \\
IRAS~15225+2350 & 12~$\pm$~1 &  0~$\pm$~1 &  5.7~$\pm$~0.6 &  0~$\pm$~1 &   4.5~$\pm$~0.6 \\
IRAS~15250+3609 & 25.6~$\pm$~0.7 &  0~$\pm$~4 & 14.5~$\pm$~0.4 &  0~$\pm$~2 &   5.4~$\pm$~0.2 \\
Arp~220 & 19~$\pm$~2 & 20~$\pm$~17 & 11.3~$\pm$~0.4 &  9~$\pm$~8 &   5.2~$\pm$~0.7 \\
FESS~J160655.82+541500.7 &  4.2~$\pm$~0.2 & 11.0~$\pm$~0.9 &  6~$\pm$~3 & 18~$\pm$~2 & 10~$\pm$~5 \\
IRAS~F16073+0209 &  6~$\pm$~1 &  0.0~$\pm$~0.4 &  7~$\pm$~2 &   0.0~$\pm$~0.6 & 11~$\pm$~3 \\
IRAS~16090-0139 & 21.3~$\pm$~0.3 &  0~$\pm$~1 &  8.2~$\pm$~0.6 &   0.0~$\pm$~0.6 &   3.7~$\pm$~0.3 \\
FESS~J161759.22+541501.3 & 11.3~$\pm$~0.4 &  0~$\pm$~1 &  0.00~$\pm$~0.01 &  0~$\pm$~1 &  0.00~$\pm$~0.01 \\
IRAS~F16156+0146 & 13~$\pm$~2 & 14~$\pm$~7 & 13.8~$\pm$~0.9 &  9~$\pm$~4 &  9~$\pm$~1 \\
IRAS~F16242+2218 & 14.9~$\pm$~0.7 & 0 &  7~$\pm$~2 & 0 &  5~$\pm$~2 \\
IRAS~F16305+4823 & 24~$\pm$~1 & 21~$\pm$~9 & 19~$\pm$~1 &  7~$\pm$~3 &   6.6~$\pm$~0.6 \\
IRAS~16300+1558 & 19~$\pm$~1 & 17~$\pm$~3 & 18.6~$\pm$~0.8 &  7~$\pm$~1 &   8.2~$\pm$~0.5 \\
IRAS~16455+4553 & 14~$\pm$~2 &  0.0~$\pm$~0.3 & 13.5~$\pm$~0.7 &   0.0~$\pm$~0.2 &  9~$\pm$~1 \\
IRAS~16468+5200W & 19.9~$\pm$~0.5 &  0.00~$\pm$~0.08 &  9.7~$\pm$~0.4 &  0.00~$\pm$~0.04 &   4.7~$\pm$~0.2 \\
IRAS~16468+5200E & 18.9~$\pm$~0.5 &  0.00~$\pm$~0.09 & 12.3~$\pm$~0.6 &  0.00~$\pm$~0.05 &   6.1~$\pm$~0.3 \\
NGC~6240 & 12.0~$\pm$~0.3 & 10.2~$\pm$~0.6 &  5.7~$\pm$~0.2 &   7.5~$\pm$~0.5 &   4.2~$\pm$~0.2 \\
IRAS~17044+6720 &  6.4~$\pm$~0.2 & 0 &  6.6~$\pm$~0.4 & 0 &   9.4~$\pm$~0.6 \\
IRAS~F17028+3616 & 11~$\pm$~1 & 38~$\pm$~8 & 11~$\pm$~2 & 24~$\pm$~6 &  7~$\pm$~2 \\
IRAS~17068+4027 & 18.2~$\pm$~0.4 &  4.8~$\pm$~0.1 & 16.1~$\pm$~0.7 &  2.38~$\pm$~0.07 &   7.9~$\pm$~0.4 \\
IRAS~17208-0014 & 15.2~$\pm$~0.3 & 16~$\pm$~1 & 13.8~$\pm$~0.4 &   8.9~$\pm$~0.8 &   7.6~$\pm$~0.3 \\
IRAS~17463+5806 & 11.5~$\pm$~0.5 & 85~$\pm$~4 & 20~$\pm$~1 & 39~$\pm$~2 &   9.1~$\pm$~0.7 \\
IRAS~17540+2935 & 19~$\pm$~1 &  0~$\pm$~3 &  3.4~$\pm$~0.7 &  0~$\pm$~2 &   1.8~$\pm$~0.4 \\
IRAS~18443+7433 & 15.3~$\pm$~0.5 &  3.7~$\pm$~0.9 & 15.5~$\pm$~0.5 &   2.2~$\pm$~0.5 &   9.0~$\pm$~0.4 \\
IRAS~18531-4616 & 22.8~$\pm$~0.5 &  1~$\pm$~3 & 16~$\pm$~2 &  1~$\pm$~1 &   6.7~$\pm$~0.7 \\
IRAS~18588+3517 & 22.4~$\pm$~0.3 & 10.4~$\pm$~0.1 &  6.2~$\pm$~0.6 &  4.32~$\pm$~0.07 &   2.6~$\pm$~0.3 \\
IRAS~20087-0308 & 15.2~$\pm$~0.6 & 0 &  5~$\pm$~1 & 0 &   3.3~$\pm$~0.7 \\
IRAS~20100-4156 & 21.9~$\pm$~0.7 &  0.0~$\pm$~0.9 &  6.4~$\pm$~0.6 &   0.0~$\pm$~0.4 &   2.8~$\pm$~0.3 \\
IRAS~20109-3003 & 20.1~$\pm$~0.4 & 17~$\pm$~1 & 21~$\pm$~1 &   7.0~$\pm$~0.6 &   8.9~$\pm$~0.4 \\
IRAS~20286+1846 & 15.7~$\pm$~0.7 &  0~$\pm$~2 & 10~$\pm$~2 &  0~$\pm$~1 &  6~$\pm$~1 \\
IRAS~20551-4250 & 15.0~$\pm$~0.4 &  8.1~$\pm$~0.2 & 15.5~$\pm$~0.6 &   4.7~$\pm$~0.2 &   8.9~$\pm$~0.4 \\
IRAS~21077+3358 & 16.5~$\pm$~0.1 & 0 &  4.1~$\pm$~0.7 & 0 &   2.4~$\pm$~0.4 \\
IRAS~21272+2514 & 17.0~$\pm$~0.3 &  0.0~$\pm$~0.8 &  7~$\pm$~1 &   0.0~$\pm$~0.5 &   3.7~$\pm$~0.6 \\
IRAS~F21329-2346 & 13~$\pm$~1 & 23~$\pm$~6 & 18.1~$\pm$~0.3 & 13~$\pm$~4 &  10.4~$\pm$~0.8 \\
IRAS~F21541-0800 &  5.7~$\pm$~0.2 & 11.8~$\pm$~0.9 & 10.3~$\pm$~0.7 & 15~$\pm$~1 &  13.1~$\pm$~0.9 \\
NGC~7172 &  5.8~$\pm$~0.2 & 13~$\pm$~1 &  9~$\pm$~1 & 17~$\pm$~1 & 11~$\pm$~1 \\
IRAS~22088-1831W & 20~$\pm$~1 & 18~$\pm$~4 & 23.0~$\pm$~0.9 &  7~$\pm$~2 &   9.3~$\pm$~0.6 \\
IRAS~22088-1831E & 21.7~$\pm$~0.4 & 19~$\pm$~1 & 21~$\pm$~1 &   7.3~$\pm$~0.6 &   8.2~$\pm$~0.4 \\
IRAS~22116+0437 & 16.8~$\pm$~0.2 & 37~$\pm$~1 & 22.0~$\pm$~0.9 &  16.4~$\pm$~0.5 &   9.7~$\pm$~0.4 \\
NGC~7479 &  8.8~$\pm$~0.2 & 16.4~$\pm$~0.3 &  5.7~$\pm$~0.5 &  14.9~$\pm$~0.4 &   5.2~$\pm$~0.5 \\
IRAS~23129+2548 & 27.0~$\pm$~0.9 &  2~$\pm$~4 & 18.1~$\pm$~0.7 &  1~$\pm$~1 &   6.2~$\pm$~0.3 \\
IRAS~F23234+0946 & 19~$\pm$~1 &  9~$\pm$~6 &  3.9~$\pm$~0.2 &  5~$\pm$~3 &   1.9~$\pm$~0.2 \\
IRAS~23230-6926 & 18.6~$\pm$~0.5 &  7.8~$\pm$~0.2 & 16.6~$\pm$~0.5 &   3.7~$\pm$~0.1 &   7.9~$\pm$~0.3 \\
IRAS~23253-5415 & 16.3~$\pm$~0.8 & 0 &  1.7~$\pm$~0.5 & 0 &   1.0~$\pm$~0.3 \\
IRAS~23365+3604 & 0 &  0~$\pm$~12 & 0 &  0~$\pm$~8 & 0 \\
\hline

\end{tabular}

\end{table*}

   \begin{figure*}[h]
	\centering
	\includegraphics[width=16cm]{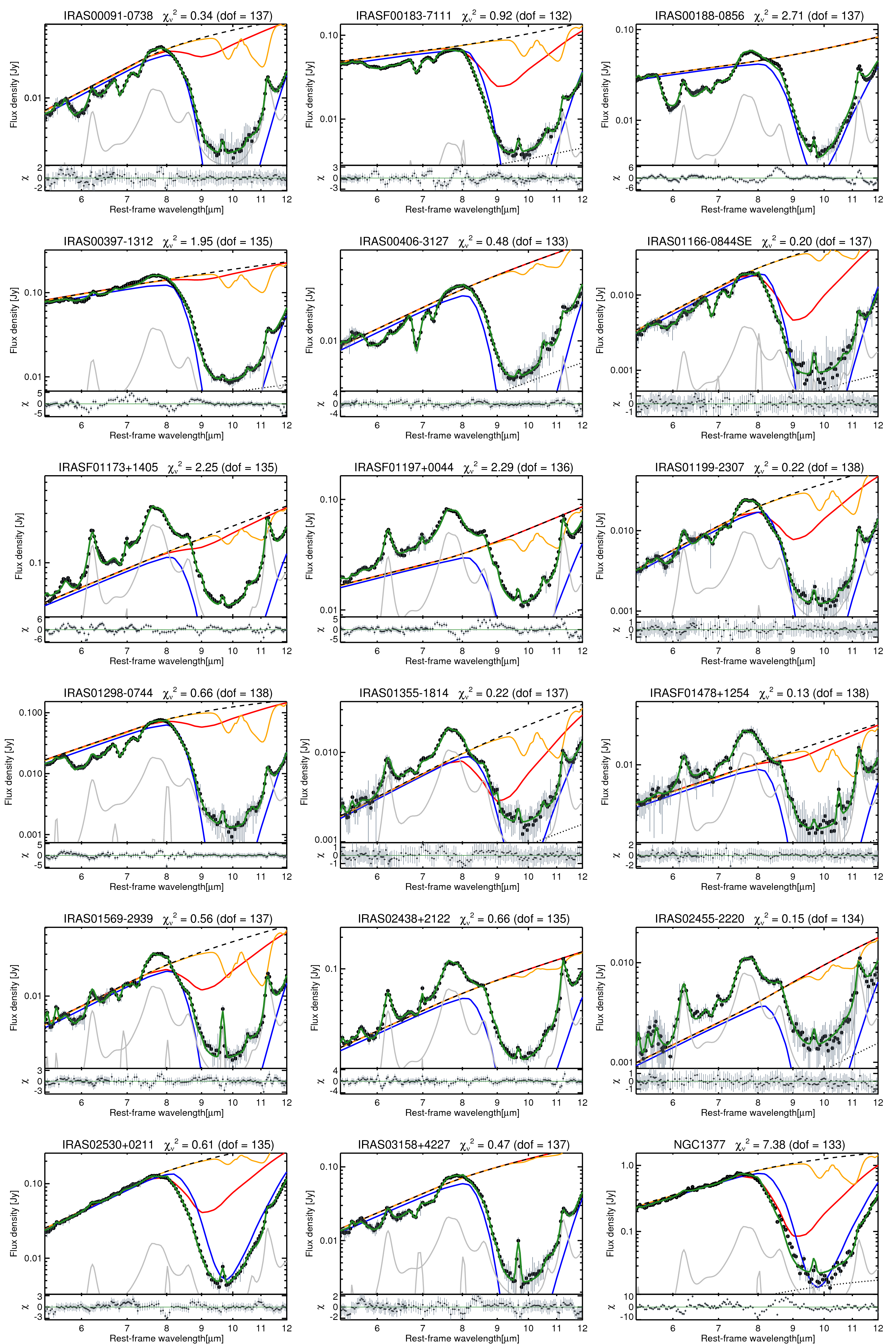}
	\caption{Results of the 5.3--12~$\mu$m spectral fitting for all spectra in our sample. The line colors and styles are the same as in the right panel of Fig. \ref{fig:fit_result}. Note that some of the components (mainly the unobscured continuum component) fall below the plotted area. }
	\label{fig:app_mineral1}%
   \end{figure*}
\setcounter{figure}{0}
   \begin{figure*}[h]
	\centering
	\includegraphics[width=16cm]{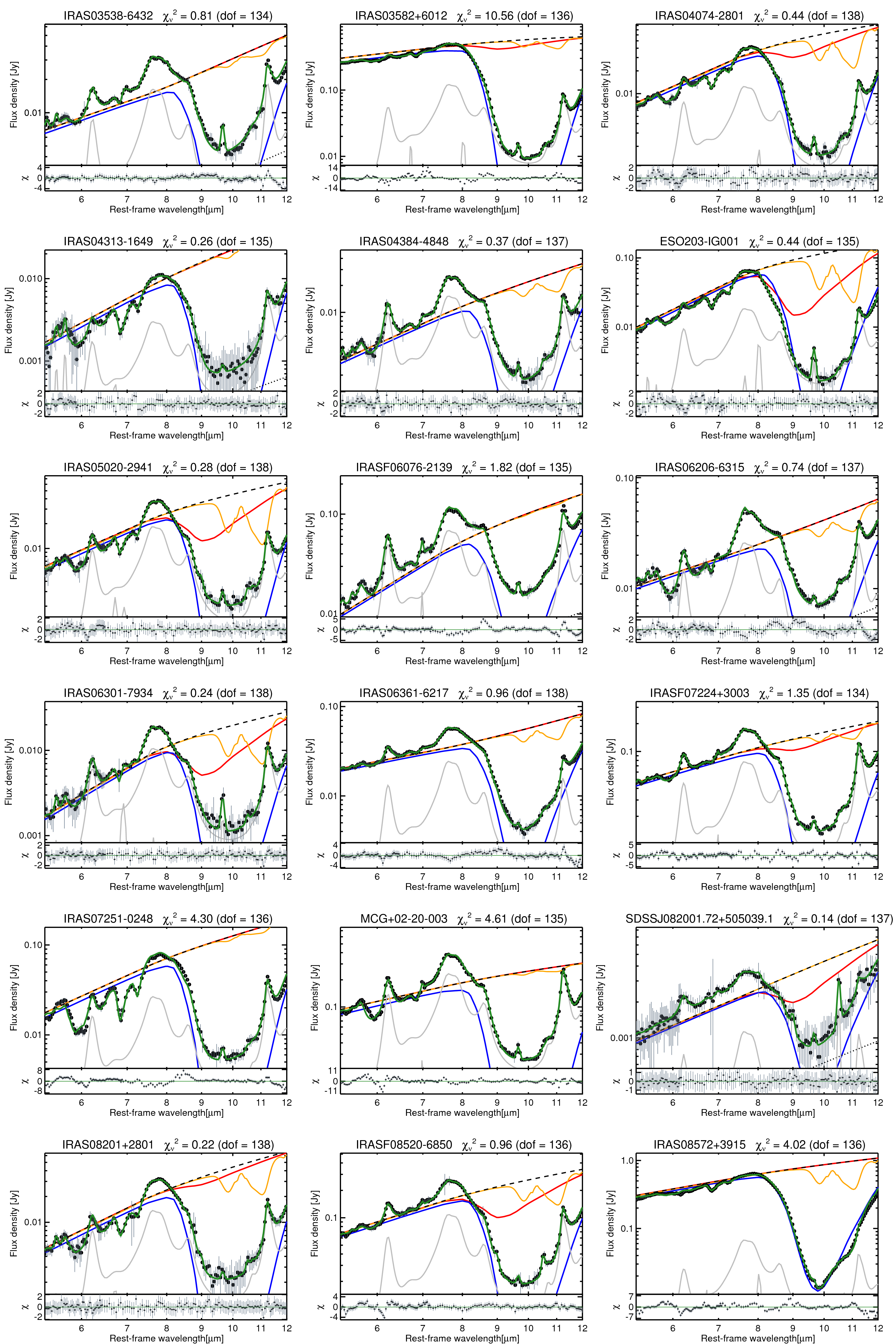}
	\caption{Continued.}
   \end{figure*}
\setcounter{figure}{0}
   \begin{figure*}[h]
	\centering
	\includegraphics[width=16cm]{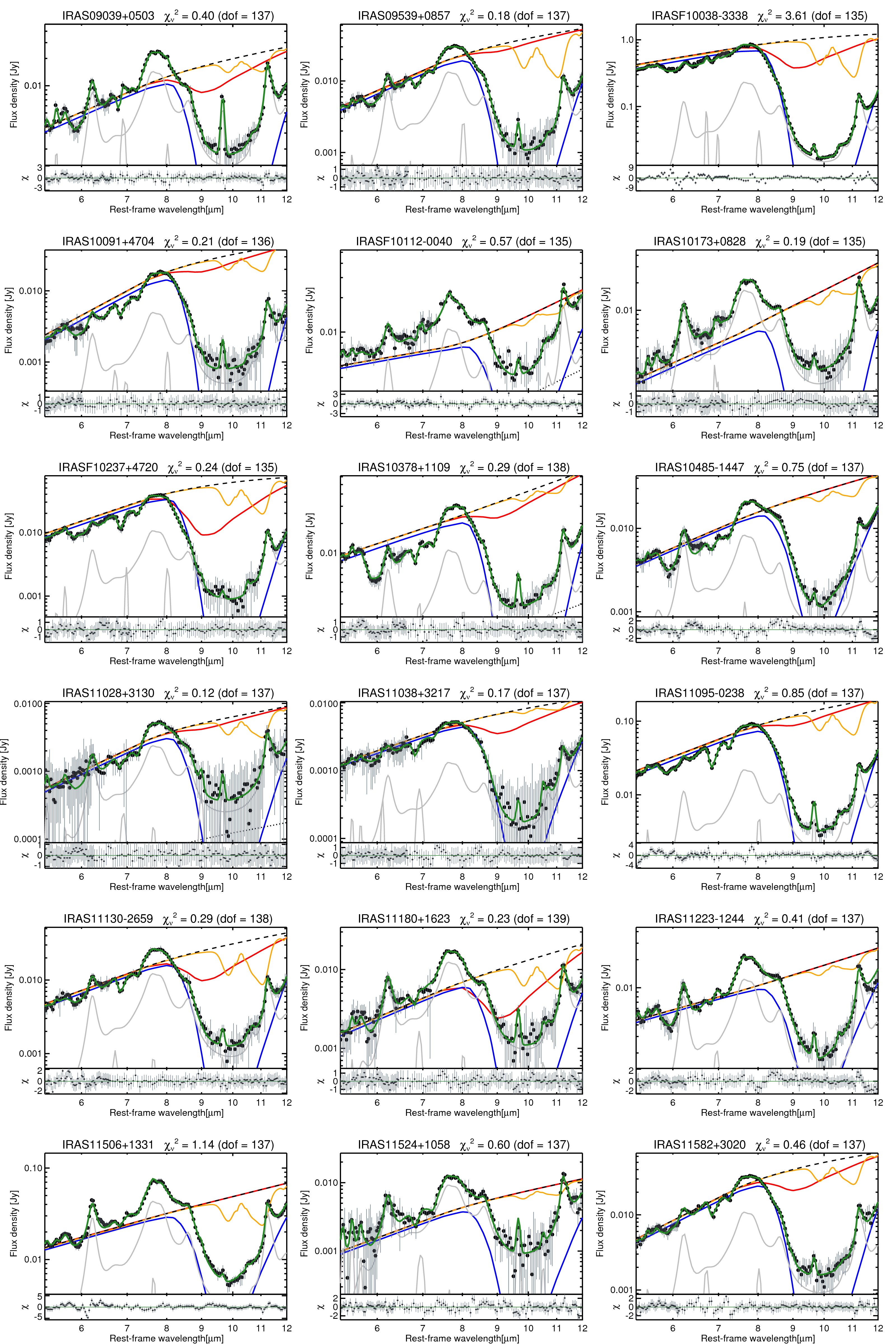}
	\caption{Continued.}
   \end{figure*}
\setcounter{figure}{0}
   \begin{figure*}[h]
	\centering
	\includegraphics[width=16cm]{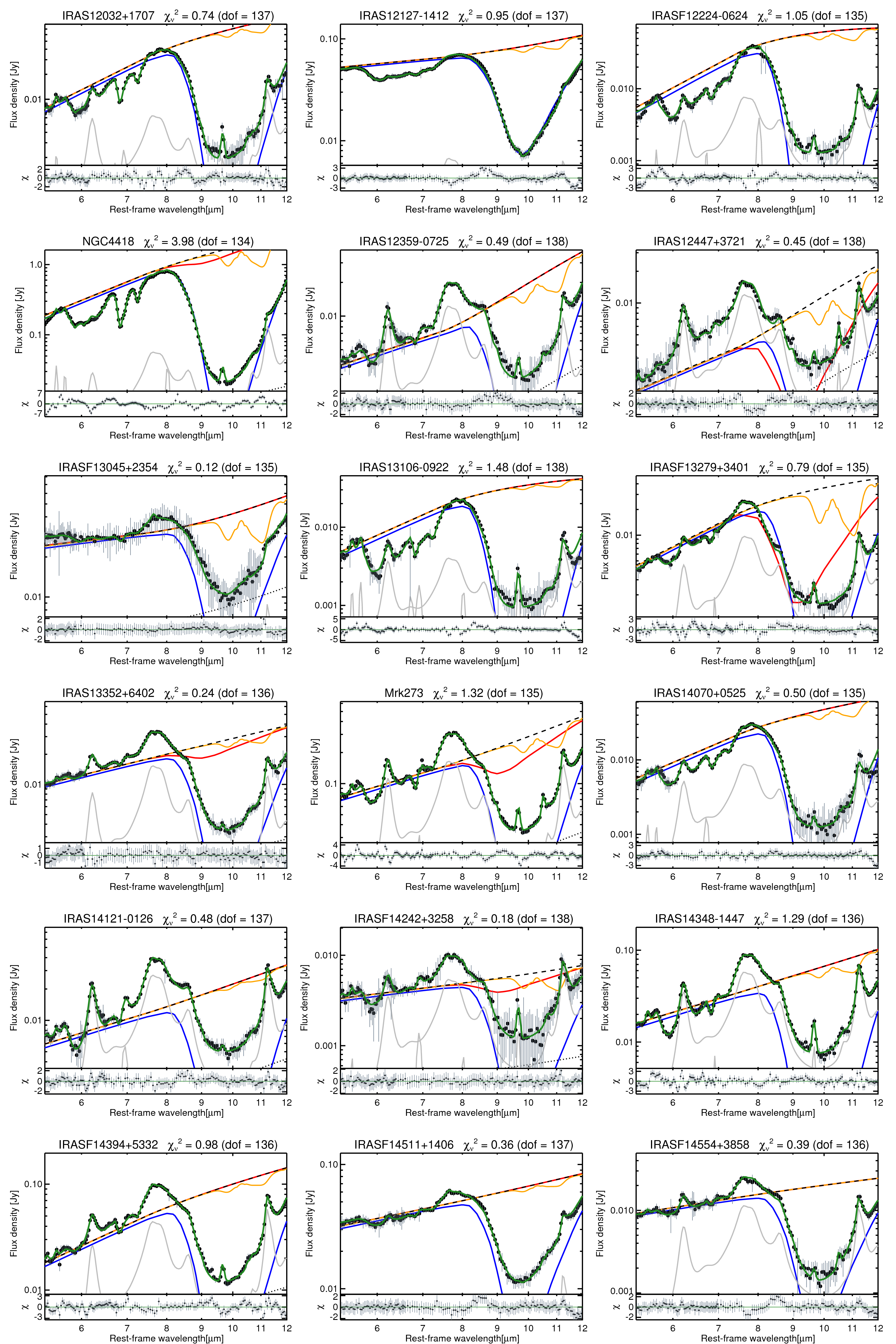}
	\caption{Continued.}
   \end{figure*}
\setcounter{figure}{0}
   \begin{figure*}[h]
	\centering
	\includegraphics[width=16cm]{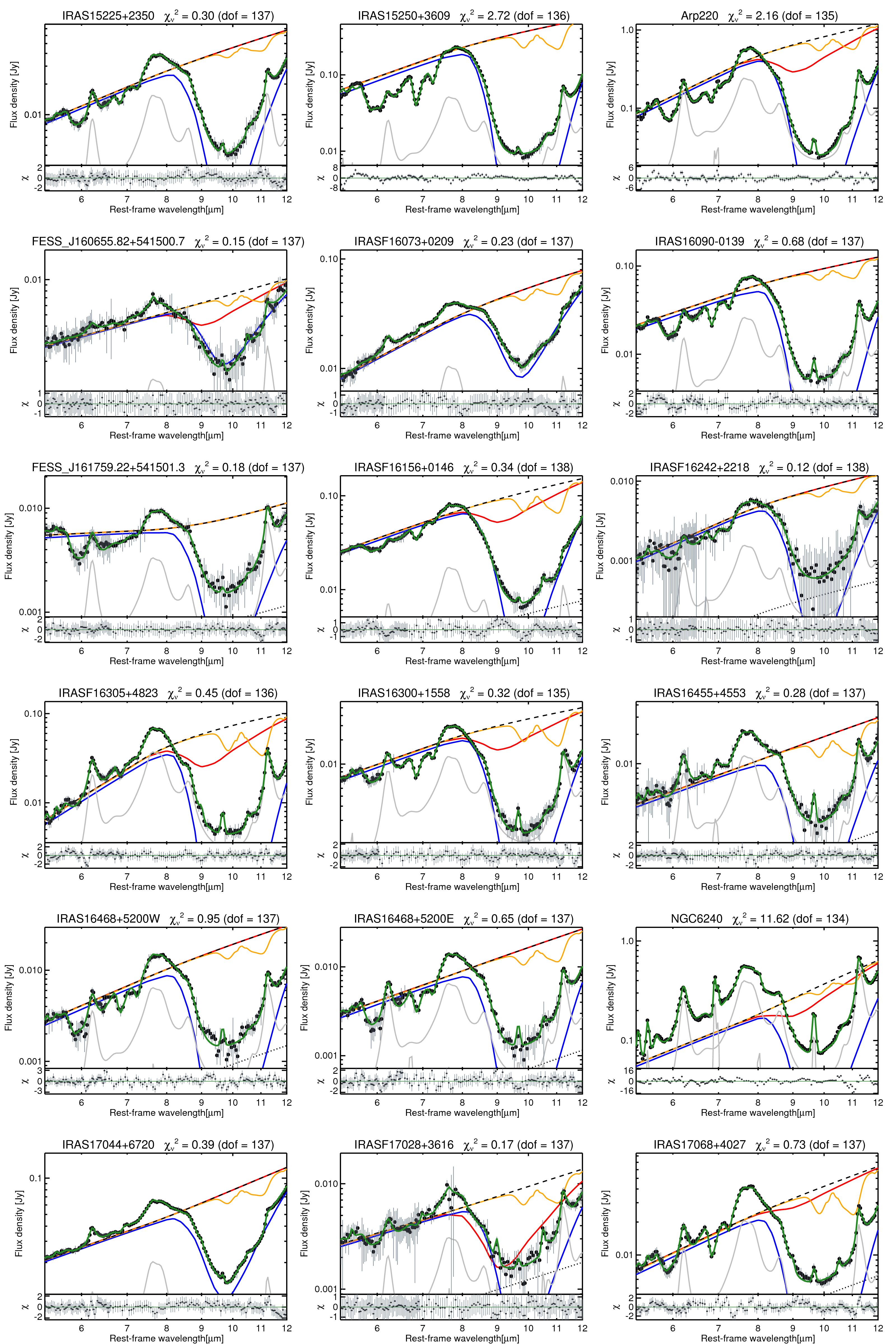}
	\caption{Continued.}
   \end{figure*}
\setcounter{figure}{0}
   \begin{figure*}[h]
	\centering
	\includegraphics[width=16cm]{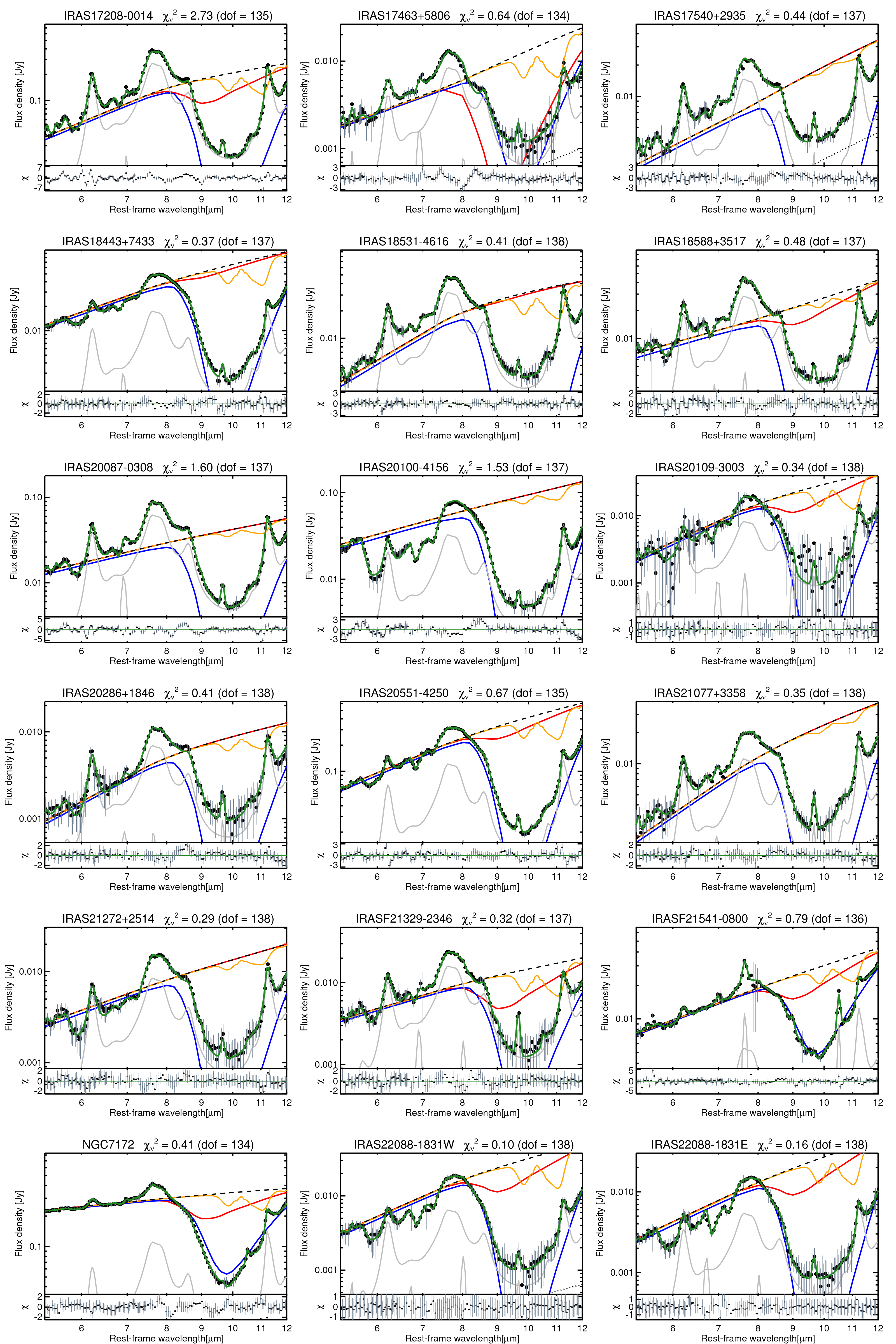}
	\caption{Continued.}
   \end{figure*}
\setcounter{figure}{0}
   \begin{figure*}[h]
	\centering
	\includegraphics[width=16cm]{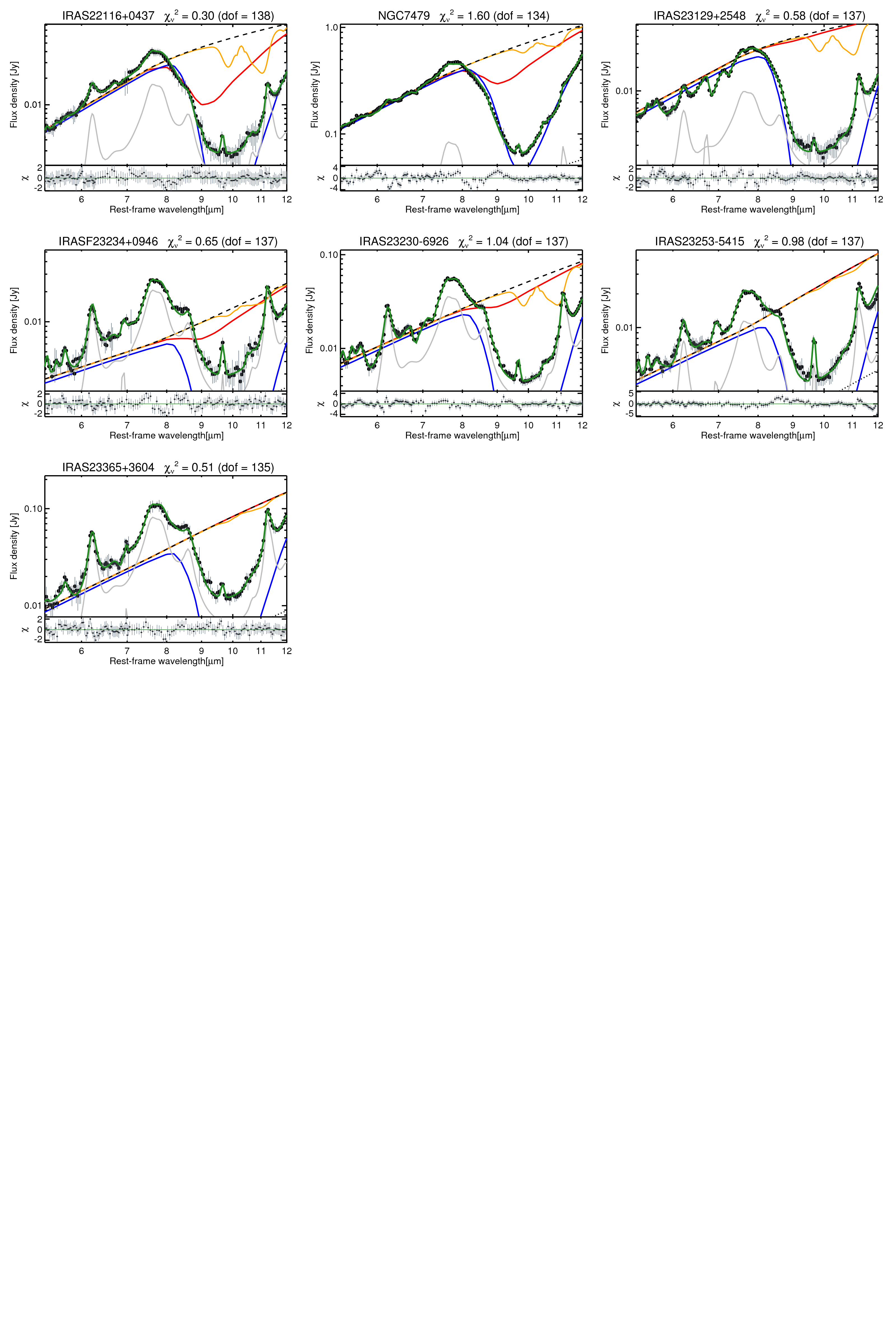}
	\caption{Continued.}
   \end{figure*}

\end{document}